\documentclass[lettersize, journal]{IEEEtran}
\usepackage{hyperref}
\usepackage{amsmath, amsfonts, amssymb}
\usepackage{algorithmic}
\usepackage{algorithm}
\usepackage{array}
\usepackage[caption=false, font=normalsize, labelfont=sf, textfont=sf]{subfig}
\usepackage{textcomp}
\usepackage{stfloats}
\usepackage{url}
\usepackage{verbatim}
\usepackage{graphicx}
\usepackage{cite}
\usepackage{booktabs}
\usepackage{multirow}
\def\vs{\emph{v.s. }}
\def\ie{\emph{i.e.}}
\def\eg{\emph{e.g.}}
\def\wrt{\emph{w.r.t. }}
\usepackage{color}
\usepackage{diagbox}
\definecolor{brown}{RGB}{139, 69, 19}
\begin{document}
	
	\title{An Effective Information Theoretic Framework for Channel Pruning}
	
	\author{
		Yihao Chen, Zefang Wang
		
		\thanks{Manuscript submitted 01 March 2023; revised 09 October 2023; accepted 15 January 2024. \textit{(Corresponding author: Yihao Chen.})}
		
		\thanks{Yihao Chen is with the College of Control Science and Engineering, Zhejiang University, China. E-mail: \textit{yihaochen@zju.edu.cn}}
				
		\thanks{Zefang Wang is with the Ningbo Innovation Center, Zhejiang University, China. E-mail: \textit{zefangwang@zju.edu.cn}}
	
		\thanks{Digital Object Identifier 10.1109/TNNLS.2024.3365194}
	}
	
	\markboth{Accepted by IEEE Transactions on Neural Networks and Learning Systems}
	{Shell \MakeLowercase{\textit{et al.}}: A Sample Article Using IEEEtran. cls for IEEE Journals}
	
	
	\maketitle

	\begin{abstract} Channel pruning is a promising method for accelerating and compressing convolutional neural networks. However, current pruning algorithms still remain unsolved problems that how to assign layer-wise pruning ratios properly and discard the least important channels with a convincing criterion. In this paper, we present a novel channel pruning approach via information theory and interpretability of neural networks. Specifically, we regard information entropy as the expected amount of information for convolutional layers. In addition, if we suppose a matrix as a system of linear equations, a higher-rank matrix represents there exist more solutions to it, which indicates more uncertainty. From the point of view of information theory, the rank can also describe the amount of information. In a neural network, considering the rank and entropy as two information indicators of convolutional layers, we propose a fusion function to reach a compromise of them, where the fusion results are defined as ``information concentration''. When pre-defining layer-wise pruning ratios, we employ the information concentration as a reference instead of heuristic and engineering tuning to provide a more interpretable solution. Moreover, we leverage Shapley values, which are a potent tool in the interpretability of neural networks, to evaluate the channel contributions and discard the least important channels for model compression while maintaining its performance. Extensive experiments demonstrate the effectiveness and promising performance of our method. For example, our method improves the accuracy by 0.21\% when reducing 45.5\% FLOPs and removing 40.3\% parameters for ResNet-56 on CIFAR-10. Moreover, our method obtains loss in Top-1/Top-5 accuracies of 0.43\%/0.11\% by reducing 41.6\% FLOPs and removing 35.0\% parameters for ResNet-50 on ImageNet. In object detection, our method yields an mAP of 37.6\% with 25.55M parameters for RetinaNet on COCO2017. 
	\end{abstract}
	
	
	\section{Introduction}
\begin{figure*}[t!]
		\centering
		\includegraphics[width=\linewidth]{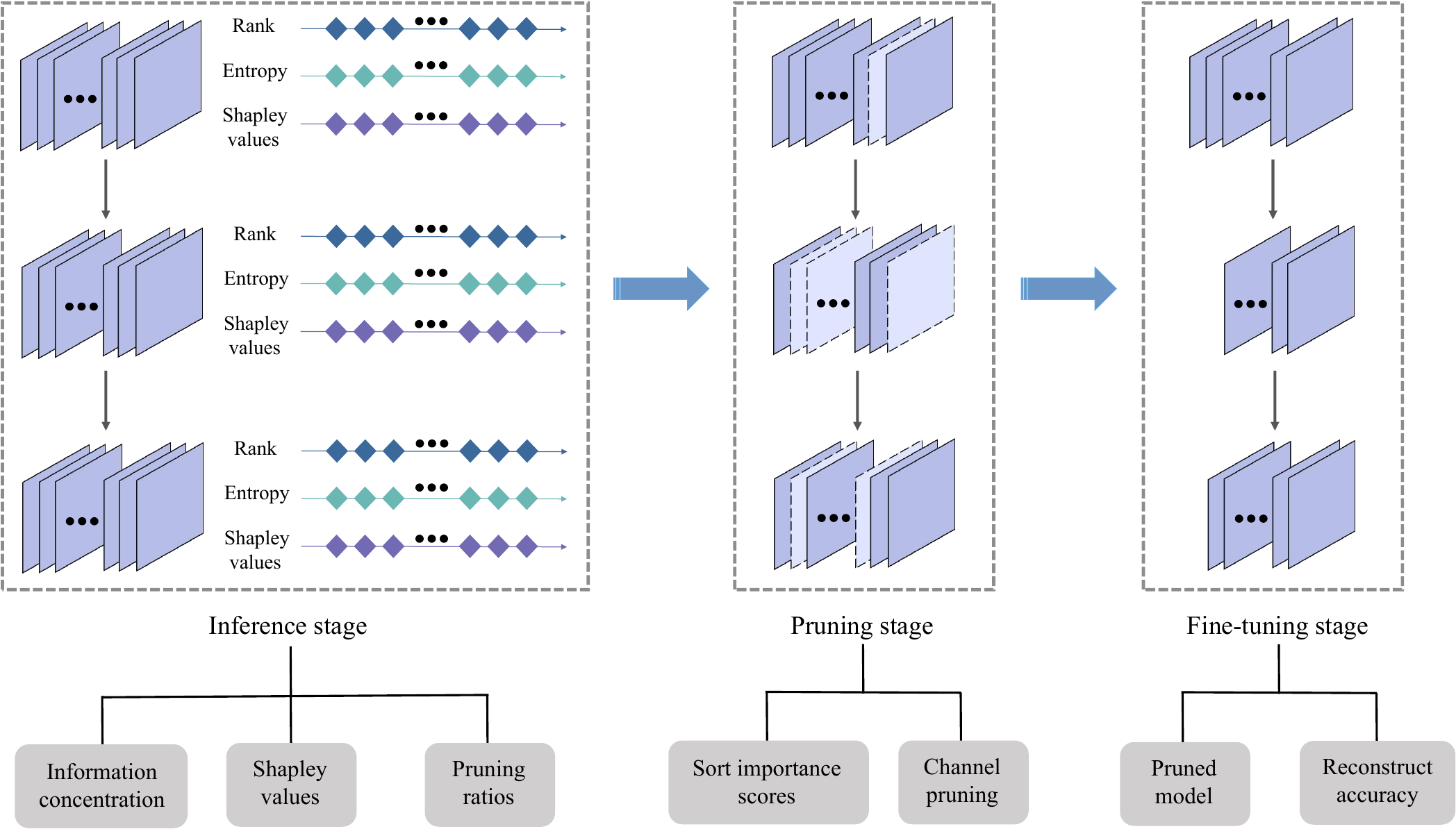}
		\caption{The overview of our proposed method. The boxes denote the channels, and the ones with dash lines indicate they are removed. In the inference stage, we feed the sampled data to the network, after that we obtain the information concentration and Shapley values. Then we assign layer-wise pruning ratios via the information concentration. In the pruning stage, we sort the importance scores of the channels represented by Shapley values and discard the least important ones in each layer. Finally, we fine-tune the pruned model again to reconstruct its accuracy.}
		\label{fig:framework}
	\end{figure*}

	\IEEEPARstart{C}{onvolutional} Neural Networks (CNNs) demonstrates its superiority in computer vision tasks such as image classification \cite{alexnet, vgg, googlenet, resnet}, object detection \cite{faster-rcnn, yolo, ssd, fpn}, semantic segmentation \cite{unet, fcn, segnet}. However, these models have a large number of parameters and a high computational cost, making them difficult to deploy on mobile and embedded devices. Even if the network is specially designed with an efficient architecture (\eg, residual connection \cite{resnet}, inception module \cite{googlenet}), over-parametrization still remains a problem. To solve this problem, a model with a small memory footprint and low computation overhead but high accuracy is required. There are two categories of network pruning that can be used to accelerate and compress a model, \ie, unstructured pruning \cite{weight-structured, weight-deep, weight-learning, weight-dynamic, weight-compression, dynamic, surgeon} and structured pruning \cite{runtime, pruning, variational, dais, carrying}. Unstructured pruning methods (\eg, weight pruning) prune the unimportant weights of the network to produce a sparse tensor. Nevertheless, the model achieved by weight pruning  can only be integrated into BLAS libraries \cite{faster} and can only be accelerated with specialized hardware \cite{eie}. On the contrary, structured pruning (\eg, filter pruning, channel pruning) methods discard the entire filters or channels in a network. As a result, since the entire filters or channels are discarded, they can achieve a lightweight network that does not require specialized software or hardware. To this end, we concentrate on channel pruning for the purpose of reducing parameters and computational costs while maintaining model accuracy, with the goal of providing a scheme for deploying the neural network in a resource-limited device. How to appropriately assign the pruning rate per layer is a difficult problem of channel pruning. Recent studies \cite{hrank, cp, thinet} empirically pre-define the pruning rate for each layer. However, they do not provide a reliable explanation for the setting of the layer-wise pruning rates. Heuristic tuning is typically required to determine how to set an appropriate pruning rate per layer \cite{emerging}. Another critical issue is how to identify unimportant channels. Previous algorithms used multiple criteria to describe the importance of the channels \cite{pfec, sfp, fpgm, hrank}, but these properties lack of interpretability when used to describe the importance of the channels in a convolutional layer. In other words, they do not provide a full explanation of the reason that the channels contribute less and are decided to be discarded. For the purpose of solving the two aforementioned issues, we present a novel method of channel pruning. Fig. \ref{fig:framework} demonstrates the framework of our method, which uses the information concentration of convolutional layers to assign layer-wise pruning ratios and channel contributions to prune the layers. During forward propagation, each layer makes different contributions to the final output of the neural network, which can be thought of as the information flowing through each layer. As a result, depending on the information that the convolutional layers contain, it is possible to assign various numbers of channels for pruning multiple layers. The information redundancy and amount in a tensor are described by the rank and information entropy (which we refer to as ``entropy'' for convenience) \cite{hrank, entropy}. We use a ``fusion'' of these two indicators to summarize how they characterize the data from various convolutional layers in the pre-trained model. The fusion values indicate the information that the convolutional layers contain, where the larger ones denote the corresponding layers are more important. According to the results, we employed the channel pruning method, selecting a larger number of channels for pruning in the less informative layers. In the pruning stage, we propose a theoretical foundation: pruning the channels that have the least contributions to the optimization of loss. This approach differs from previous methods, such as magnitude-based pruning, where channels are removed based on their importance scores \cite{hrank, sfp, pfec, entropy}. Shapley values are a potent tool in interpretability of deep learning \cite{shapley, cicc}, and are naturally suited for assessing the channel contributions because they can fairly distribute the \textit{average marginal contributions} to them and explicitly model the importance of the channels with feature attribution explanation. We also calculate the Shapley values along with the rank and entropy of the layers. Pruning has a less adverse effect on the performance of a model for the channels with the lowest Shapley values since they have fewer contributions to optimization. For the purpose of reconstructing the accuracy of the pruned model, we re-train the network again.
	
	\textbf{Contributions:} The following is a summary of our contributions: (1) For the purpose of obtaining an overall indicator of the information concentration of convolutional layers, we propose a fusion function that reaches a compromise of the information indicated by rank and entropy. According to the fusion values, we determine the layer-wise pruning rates. (2) We propose that an appropriate pruning criterion can be the contributions to the optimization of loss and recommend using Shapley values as a potent tool with the interpretability of CNNs to assess the channel contributions. (3) Extensive experiments for pruning architectures on CIFAR-10 \cite{cifar} and ImageNet \cite{imagenet} using VGGNet \cite{vgg}, ResNet \cite{resnet}, and DenseNet \cite{densenet} for image classification, and COCO2017 \cite{coco} using RetinaNet \cite{retinanet}, FSAF \cite{fsaf}, ATSS \cite{atss}, and PAA \cite{paa} for object detection, demonstrate the effectiveness and superiority of our method.
	
	\section{Related Works}
	
	\textbf{Pruning criteria:} Previous methods in terms of structured pruning prune a model at multiple levels (\eg, filter and channel), where channel pruning correlates to filter pruning \cite{non-structured, filter-in-filter}. The capacity and complexity of a model are reduced due to discarding the filters or channels, but this inevitably degrades the model's accuracy. Therefore, it is generally accepted that removing the least important filters or channels will minimize the loss in accuracy. For the purpose of achieving this target, the question of how to recognize the unimportant filters or channels is raised. Prior study \cite{taylor} approximates the change in the loss function to determine the importance of the filters, then removes the least important ones. The importance of the filters is estimated using a variety of criteria, including the geometric median \cite{fpgm}, $\ell_1$-norm \cite{pfec}, $\ell_2$-norm \cite{sfp}, rank \cite{hrank}, and entropy \cite{entropy}. The Average Percentage of Zeros (APoZ) is used as the importance score for the channel-level pruning method \cite{apoz}, which determines each channel's sparsity based on it. In addition, CP \cite{cp} discards the filters based on LASSO regression that produce the less informative channels along with the corresponding channels in the next layer, and ThiNet \cite{thinet} removes a channel underlying the statistics information calculated from the next layer. The most recent works \cite{fisher, fisher-transformer} also make use of fisher information as an importance scoring metric and CHEX \cite{chex} uses column subset selection to remove the channels that are not important. White-Box \cite{white-box} and MFP \cite{MFP} empirically assign layer-wise pruning ratios using class contribution and geometric distance, respectively. CLR-RNF \cite{CLR-RNF} employs a computation-aware measurement to determine pruning ratios, pruning more weights in less computation-intensive layers. CATRO \cite{CATRO} pre-defines layer-wise pruning ratios using class information from a few samples to measure the joint impact of multiple channels. ABP \cite{ABP} discards fewer filters in significant layers and more in less significant ones, after examining each layer's training importance. The MDP\cite{mdp} framework enables multi-dimensional pruning in convolutional neural networks. The CLFIP\cite{clfip} method incorporates classification loss and feature importance into the layer-wise channel pruning process. PCP\cite{pcp} iteratively prunes channels and estimates the subsequent accuracy drop. For point cloud neural networks, the JP\cite{jp} framework reduces redundancies along multiple dimensions. The 3D-P\cite{3d-p} framework prunes based on the importance of each frame or point for 3D action recognition.
	
	\textbf{Pruning rate:} Recent research assigns the layer-wise pruning rates for convolutional layers in accordance with various rules when the model is enforced with a filter or channel sparsity. The pruning rate for the convolutional layers is pre-defined in rule-based methods, demonstrating that we are aware of the proportion (number) of filters or channels which will be discarded beforehand \cite{emerging}. Early works \cite{entropy, sfp, fpgm, actd} adopt a constant percentage to discard the filters or channels in each layer. On the contrary, HRank \cite{hrank}, ThiNet \cite{thinet} and CP \cite{cp} empirically assign various layer-wise pruning ratios. In addition, NISP \cite{nisp} pre-defines the layer-wise pruning ratios based on a variety of requirements, including FLOPs, memory and accuracy. After examining the sensitivity of each convolutional layer, CC \cite{cc} and PFEC \cite{pfec} discard fewer filters in the sensitive layers and remove more filters in the insusceptible layers. For the purpose of calculating the layer-wise pruning ratios, SPP \cite{spp} performs Principal Component Analysis (PCA) on each layer and uses the reconstruction error to evaluate the sensitivity. Network Slimming \cite{slimming} first trains with channel-level sparsity-induced regularization before establishing a global pruning rate to discard the channels for the entire network. In addition, AMC \cite{amc} uses DDPG \cite{ddpg} introduced in Reinforcement Learning to learn a precise pruning ratio per layer.
	
	\textbf{Pruning schedule:} A network can be pruned using a variety of pruning schedules, which can generally be divided into three categories. (1) One-shot \cite{pfec, emerging}: remove the filters or channels of multiple convolutional layers in the network at once. NISP \cite{nisp} prunes the CNNs by removing the least important neurons and then fine-tuning the compact network to preserve its predictive power. PFEC \cite{pfec} and CC \cite{cc} prune the layers at once and further fine-tune the pruned network. Different from PFEC \cite{pfec} which prunes the unimportant weights according to $\ell_1$-norm only, GReg \cite{npgr} first imposes a $\ell_2$-norm penalty to drive the unimportant weights to zero and prune those with the least $\ell_1$-norm until the penalty factor for the weight is updated to reach a pre-set ceiling. By using generative adversarial learning, GAL \cite{gal} prunes the filters along with other structures, such as channels, branches, and blocks, in an end-to-end and label-free paradigm. (2) Progressive \cite{emerging}: the network under a simultaneous training and pruning process, and the network sparsity gradually goes from zero until the target number. The method \cite{taylor} alternates iterations of pruning the least important neuron and fine-tuning until the the pre-determined desired trade-off between accuracy and pruning objective is reached. SFP \cite{sfp} and FPGM \cite{fpgm} prune the filters based on their importance at the end of each training epoch. DSA calculates the gradient-based optimization to discover the layer-wise pruning ratios and prunes the network when training the network from scratch. Prior work \cite{fisher} discards the least important channel and the corresponding coupled ones at every several iterations until the FLOPs reduces to the desired amount. (3) Iterative \cite{pfec, emerging}: discard the unimportant filters or channels and fine-tune the pruned network, then the process is repeated until the model achieves the desired sparsity. Previous studies \cite{entropy, thinet} prune the pre-trained network and fine-tune the pruned model layer by layer with one or two epochs, and finally re-train the model with more additional epochs when all the layers are pruned. Without retraining, HRank \cite{hrank} fine-tunes the pruned network in a layer-by-layer fashion. An iterative two-step process is used by CP \cite{cp} to prune each layer, and after pruning the network is fine-tuned. Additionally, there are works \cite{acp, amc} which prune and fine-tune the network repeatedly up to the pre-determined number of iterations, and the pruned network is fine-tuned after thorough pruning. SCL \cite{SCL} applies PCA to each layer, using network connections as regularization. It initially trains with sparsity-induced regularization, then sets a global pruning rate, and finally uses reinforcement learning to optimize layer-specific pruning ratios.
	
	\textbf{Discussion:} In contrast to unstructured pruning, the model produced by structured pruning algorithms can be easily integrated into BLAS libraries and deployed to resource-limited devices. As far as we know, our method, focusing on channel pruning, is the first to assign layer-wise pruning ratios for convolutional layer pruning based on the rank and entropy for the convolutional layers in the unpruned model, which offers a more reliable method on the setting of the layer-wise pruning ratios. In addition, our method uses Shapley values to explain the contributions of the channels rather than importance-based criteria, and it discards the channels with the fewest contributions to accelerate the model during training and inference while retaining accuracy.
	
	\section{Methodology}
	
	\subsection{Problem Formulation} In a CNN-based model, for the $i_{th}$ convolutional layer $C_i$ where $1 \leq i \leq L$ and $L$ denotes the total number of convolutional layers, we assume that the number of input channels and output channels are represented by $c_i$ and $c_{i+1}$, respectively. In $C_i$, the height and weight of the feature maps are denoted as $h_i$ and $w_i$, respectively. A set of filters $K_i = \{K_i^1, K_i^2, \ldots, K_i^{c_i}\} \in \mathbb{R}^{c_i \times c_{i-1} \times k_i \times k_i}$ is in the $i_{th}$ convolutional layer, where $K_i^j \in \mathbb{R}^{c_{i-1} \times k_i \times k_i}$ represents the $j_{th}$ filter of the $i_{th}$ convolutional layer, $j \in \{1, 2, \ldots, c_i\}$ and $k_i$ denotes the kernel size of the filter. Applying $c_{i + 1}$ filters on $c_i$ channels transforms the input feature maps $M_i \in \mathbb{R}^{c_i \times h_i \times w_i}$ into output feature maps $M_{i+1} \in \mathbb{R}^{c_{i + 1} \times h_{i + 1} \times w_{i + 1}}$. In the convolutional layers, $c_i$ channels are split into two groups, \ie, the discarded channels $Q_i = \{C_i^{Q_i^1}, C_i^{Q_i^2}, \ldots, C_i^{Q_i^{q_i}}\}$ and the remaining channels $U_i = \{C_i^{U_i^1}, C_i^{U_i^2}, \ldots, C_i^{U_i^{u_i}}\}$, where $q_i$ and $u_i$ are the number of discarded and remaining channels, respectively. For $Q_i$, $U_i$, $q_i$ and $u_i$, we have: 
	\begin{equation}  
		\left\{  
		\begin{array}{lr} Q_i \cup U_i = C_i, &  \\  
			Q_i \cap U_i = \varnothing, & \\
			q_i + u_i = c_i.
		\end{array}  
		\right.  
	\end{equation} Let an indicator function $I_i^j$ denote whether the $j_{th}$ channel in the $i_{th}$ convolutional layer is pruned, then we have:
	\begin{equation} I_i^j=
		\begin{cases}
			0,& C_i^j \in Q_i, \\
			1,& C_i^j \in U_i.
		\end{cases}
	\end{equation} Assume that $o (C_i^j)$ represents the importance of the $j_{th}$ channel in the $i_{th}$ convolutional layer, and we sort $o (C_i^j)$ to prune $u_i$ unimportant channels. Therefore, channel pruning can be defined as the optimization problem as below:
	\begin{equation}
		\label{eq:opt}
		\begin{split}
			&\min\limits_{I_i^j} \sum\limits_{i=1}^{L} \sum\limits_{j=1}^{c_i} I_i^j o(C_i^j),
			\\&
			s. t. \; \sum_{j=1}^{c_i} I_i^j = q_{i}.
		\end{split}
	\end{equation} which demonstrates that our target is to discover the least important channels for the purpose of minimizing the information of the removed channels.
	
	\subsection{Information Concentration}
	\label{sec:concentration} 
\begin{figure*}[ht!]
		\centering
		
		\subfloat[Rank - ResNet-20]{
			\begin{minipage}[t]{0.23\textwidth}
				\centering
				\includegraphics[scale=0.25]{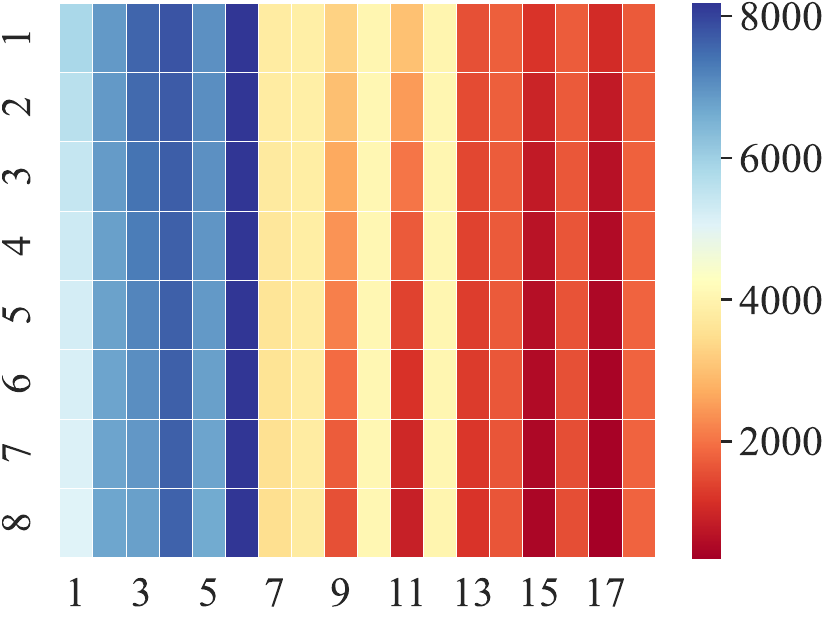}
				\label{fig:rank_entropy_heatmap_resnet20_rank}
			\end{minipage}
		}
		\subfloat[Rank - ResNet-32]{
			\begin{minipage}[t]{0.23\textwidth}
				\centering
				\includegraphics[scale=0.25]{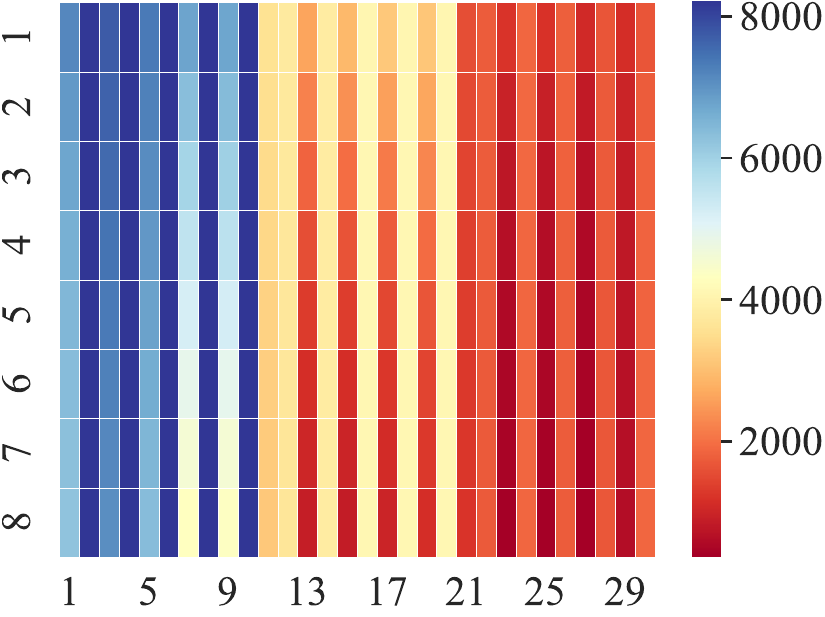}
				\label{fig:rank_entropy_heatmap_resnet32_rank}
			\end{minipage}
		}
		\subfloat[Rank - ResNet-18]{
			\begin{minipage}[t]{0.23\textwidth}
				\centering
				\includegraphics[scale=0.25]{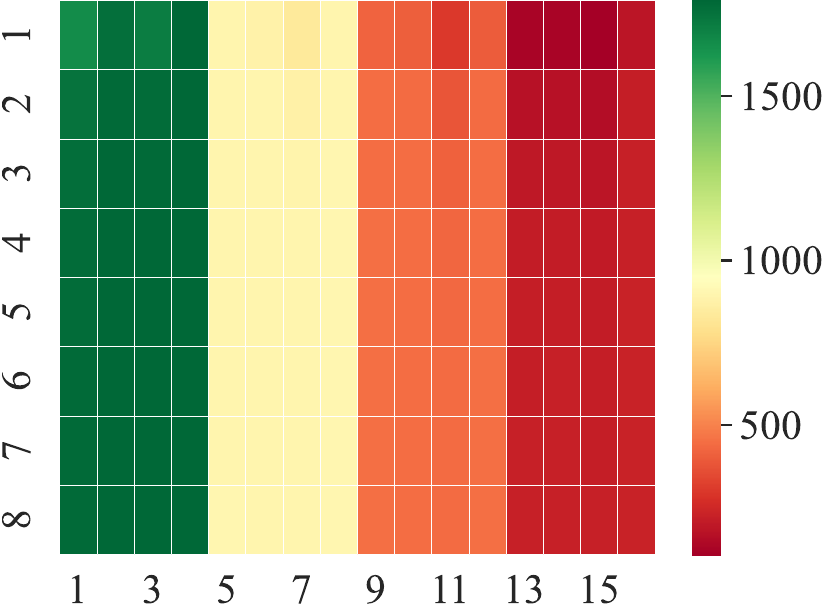}
				\label{fig:rank_entropy_heatmap_resnet18_rank}
			\end{minipage}
		}
		\subfloat[Rank - ResNet-34]{
			\begin{minipage}[t]{0.23\textwidth}
				\centering
				\includegraphics[scale=0.25]{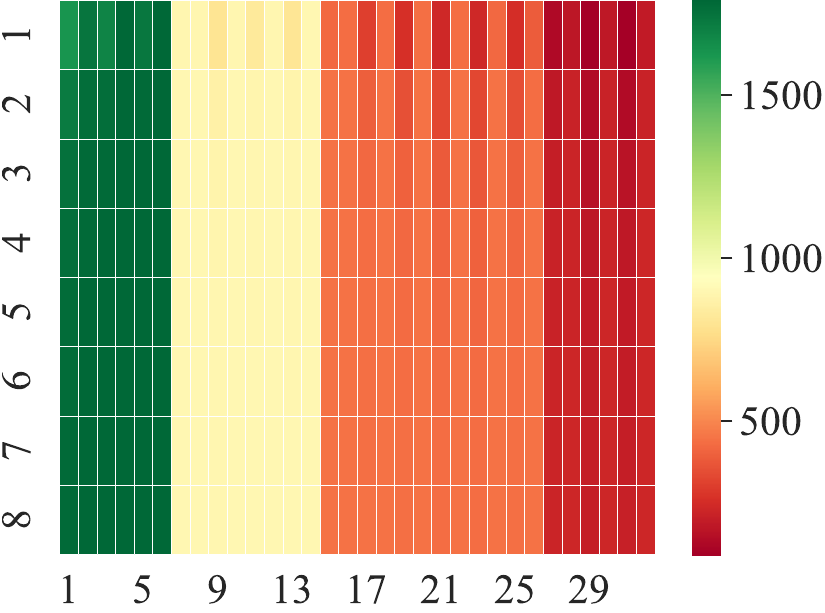}
				\label{fig:rank_entropy_heatmap_resnet34_rank}
			\end{minipage}
		}
		
		\subfloat[Entropy - ResNet-20]{
			\begin{minipage}[t]{0.23\textwidth}
				\centering
				\includegraphics[scale=0.25]{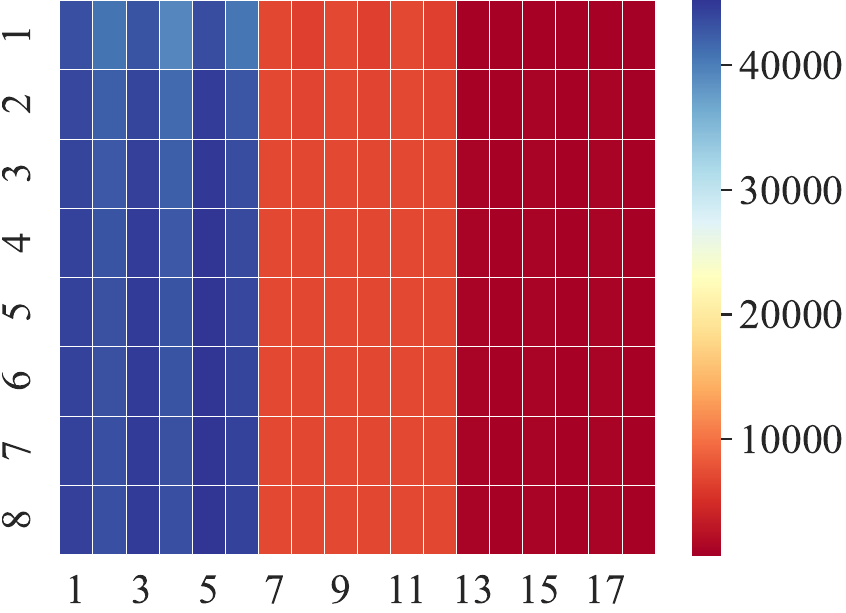}
				\label{fig:rank_entropy_heatmap_resnet20_entropy}
			\end{minipage}
		}
		\subfloat[Entropy - ResNet-32]{
			\begin{minipage}[t]{0.23\textwidth}
				\centering
				\includegraphics[scale=0.25]{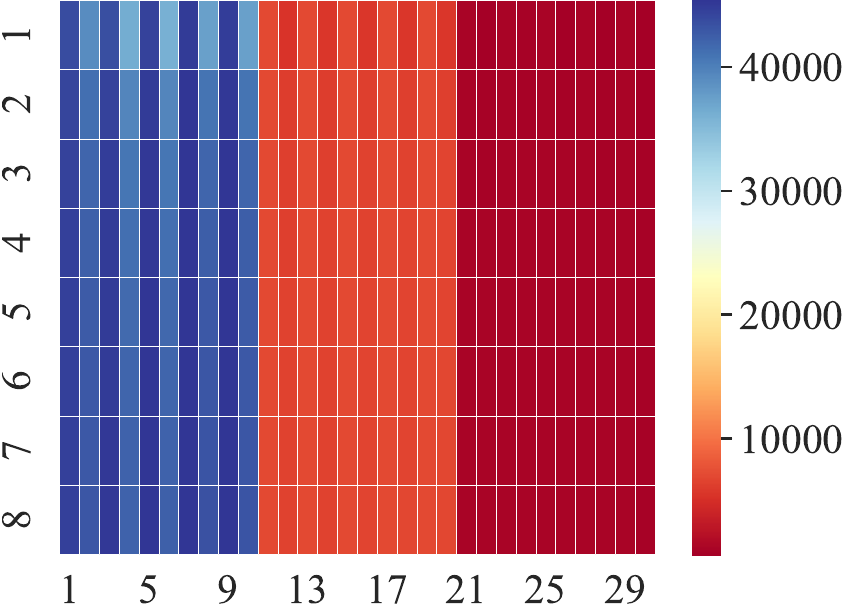}
				\label{fig:rank_entropy_heatmap_resnet32_entropy}
			\end{minipage}
		}
		\subfloat[Entropy - ResNet-18]{
			\begin{minipage}[t]{0.23\textwidth}
				\centering
				\includegraphics[scale=0.25]{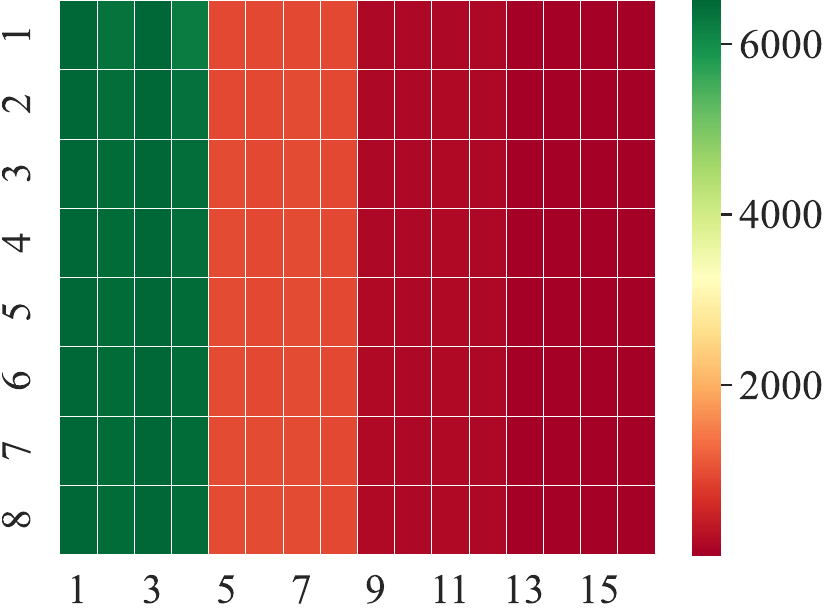}
				\label{fig:rank_entropy_heatmap_resnet18_entropy}
			\end{minipage}
		}
		\subfloat[Entropy - ResNet-34]{
			\begin{minipage}[t]{0.23\textwidth}
				\centering
				\includegraphics[scale=0.25]{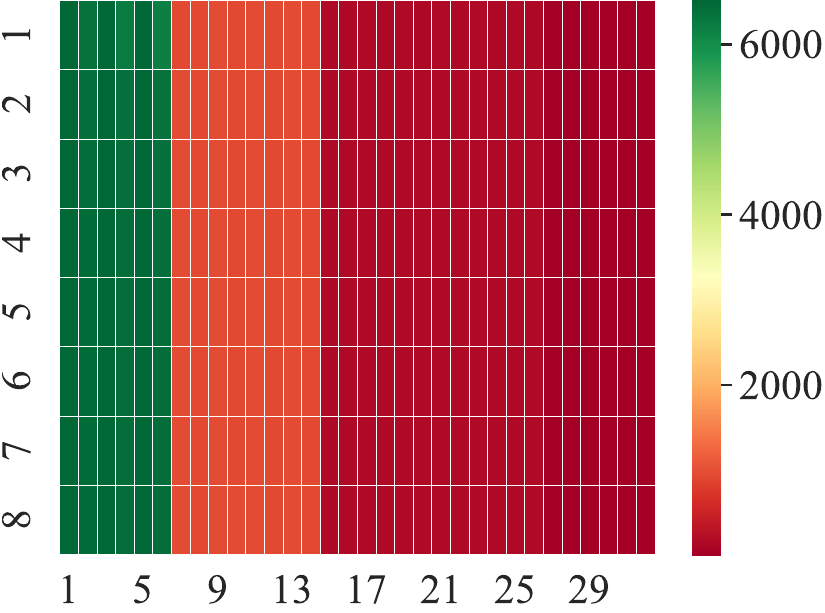}
				\label{fig:rank_entropy_heatmap_resnet34_entropy}
			\end{minipage}
		}
		
		\caption{Average statistics of rank and entropy for convolutional layer outputs under various input image batches. The convolutional layer indices are shown on the x-axis for each sub-figure, while the number of image batches are shown on the y-axis. The batch size of images for ResNet-20 and ResNet-32 is 256, while for ResNet-18 and ResNet-34 is 32. The sub-figures show that the rank and entropy for the convolutional layer outputs (the columns of each sub-figure) remain unchanged irrespective of image batches.}
		\label{fig:rank_entropy_heatmap}
	\end{figure*}

Previous works \cite{sfp, fpgm, thinet, actd} remove the same proportions of filters in the convolutional layers, and the works \cite{hrank, cp} empirically assign layer-wise pruning ratios. Different from them, we intend to assign layer-wise pruning ratios with information theory. We start by investigating the rank and entropy of the convolutional layer outputs in the pre-trained model to identify the information redundancy and amount that is contained in a layer. Then, we obtain the multiplication of them as a ``fusion'', which can be seen as the information concentration of the layers, taking into account the rank and entropy as two information indicators of convolutional layer outputs. If a convolutional layer is less informative, we can assign a larger number to prune the channels in the corresponding layer. A low-rank tensor typically contains lots of \textit{redundant information}. If we suppose a matrix as a system of linear equations, a higher-rank matrix represents there exist more solutions to it, which indicates more uncertainty. From the point of view of information theory, the rank can describe the amount of information. In the case of channel pruning, a higher-rank tensor of the channel in a convolutional layer is more informative than the lower-rank one \cite{hrank}. Additionally, a layer with a low average rank per channel suggests that multiple channels in it probably contain repetitive information, making it possible to compress it into a more compact layer with negligible loss in important information. Identifying the importance of features via rank provides a way to identify the redundant features with interpretability. To accurately calculate the rank of a matrix, we use the Singular Value Decomposition (SVD) method. First, we decompose the matrix A using SVD, yielding orthogonal matrices $U$,$\Sigma $, and $V^T$. Then, we count the number of non-zero singular values in the diagonal matrix$ \Sigma$, which is the rank of matrix A. Compared to directly calculating the matrix rank, this method avoids the NP-hard problem and can obtain a relatively accurate rank estimate. The key insight in SVD is that the number of non-zero elements in the $\Sigma$ matrix is equivalent to the rank of the matrix $A$. Hence, we can compute the rank of the matrix $A$ by performing SVD decomposition and counting the number of non-zero elements in the $\Sigma$ matrix. Previous work \cite{hrank} demonstrated the rank of each convolutional layer almost remains the same under various image batches. Inspired by the conclusion, we discover that a small portion of images can generate the outputs to estimate the rank for convolutional layers, as demonstrated in Figs. \ref{fig:rank_entropy_heatmap_resnet20_rank} $\sim$  \ref{fig:rank_entropy_heatmap_resnet34_rank}. Therefore, in the dataset with $N$ images, we first randomly fetch out $B$ images and feed them to the network for the sum of rank for the outputs per layer. Then, for the $i_{th}$ convolutional layer $C_i$, we can get the average rank per channel as:
	
	\begin{equation}
		\label{eq:avg_rank} R(C_i) = \frac{\sum\limits_{b=1}^{B} \sum\limits_{j=1}^{c_i} Rank (K_i^j(b, j, :, :))}{c_i}
	\end{equation} Generally, entropy measures the disorder of a system or the uncertainty of an event. In information theory, entropy can be regarded as the \textit{expected amount of information} \cite{information_theory}. A low-entropy convolutional layer suggests that the channels inside it contain less information in the channel pruning scenario. Thus, we could remove more channels in such convolutional layers. Entropy is naturally fit for selecting features by the amount of information they contain, therefore, it provides a more interpretable solution for identifying less informative convolutional layers. Additionally, we find that similar to the calculation for the rank of convolutional layers, the estimation of entropy for the convolutional layers may be generated by feeding only one or more batches of images, as shown in Figs. \ref{fig:rank_entropy_heatmap_resnet20_entropy} $\sim$ \ref{fig:rank_entropy_heatmap_resnet34_entropy}. Given $B$ input images, we first use the softmax function to map $C_i^j$ between $0$ and $1$, after which the outputs can be regarded as the probability distribution for the channels in $C_i$:
	\begin{equation} p(C_i^j) = Softmax(C_i^j) = \frac {\sum\limits_{b=1}^{B} e^{K_i^j(b, :, :, :)}} {\sum\limits_{b=1}^{B} \sum\limits_{j=1}^{{c_i}} e^{K_i^j(b, j, :, :)}}.
	\end{equation} After obtaining the probability distribution of the channels in a convolutional layer, we can get the average entropy per channel of $C_i$ as:
	\begin{equation}
		\label{eq:avg_entropy} E(C_i) = - \frac{\sum\limits_{j=1}^{{c_i}} p(C_i^j) \log {p(C_i^j)}}{c_i}
	\end{equation}
	
	 As an off-line analysis, we first obtain the average statistics of rank and entropy per channel of convolutional layer outputs via Eqn. \ref{eq:avg_rank} and Eqn. \ref{eq:avg_entropy} on the pre-trained model. They can be considered as two information indicators of the convolutional layers, which measure the information from different aspects. The sub-figures in Fig. \ref{fig:rank_entropy_heatmap} illustrate that under various image batches, the rank and entropy for convolutional layer outputs nearly remain unchanged with minor fluctuations. Additionally, the internal variations in rank and entropy are not entirely consistent. Therefore, we normalize them to $[l, m]$ and propose a fusion function that reaches a compromise of these two indicators to obtain an overall indicator as the information concentration of convolutional layers, reducing the inconsistencies and employing the information complementarity between these two indicators:
	\begin{equation}
		\label{eq:fusion} O(C_i) = \prod \limits_{X, Y} ((m - l) \frac {X - \min{Y}}{\max{Y} - \min{Y}} + l),
	\end{equation} where $X$ represents $R(C_i)$ and $E(C_i)$, and $Y$ represents $\{R(C_i)\} $ and $\{E(C_i)\}$ $\forall i \in \{1, 2, \ldots, L\}$, respectively. After obtaining $O(C_i)$, we also normalize it to $[l, m]$. The overall indicator offers a trustworthy representation of the information that the channels in the convolutional layers contain since it includes rank and entropy characterizations. Though rank and entropy can separately provide solutions with interpretability for uninformative layers, combining them eliminates some potential errors they themselves may have partially caused. After obtaining the fusion value, we assign $u_i$ channels to prune the $i_{th}$ convolutional layer. Since we consider the layers with lower fusion values as being less informative, $u_i$ in such layers should be set to a higher value. Otherwise, we prune fewer channels in the layers.
	
	\begin{figure*}[ht!]
		\centering
		
		
		\subfloat[Rank - ResNet-32]{
			\begin{minipage}[t]{0.3\textwidth}
				\centering
				\includegraphics[scale=0.3]{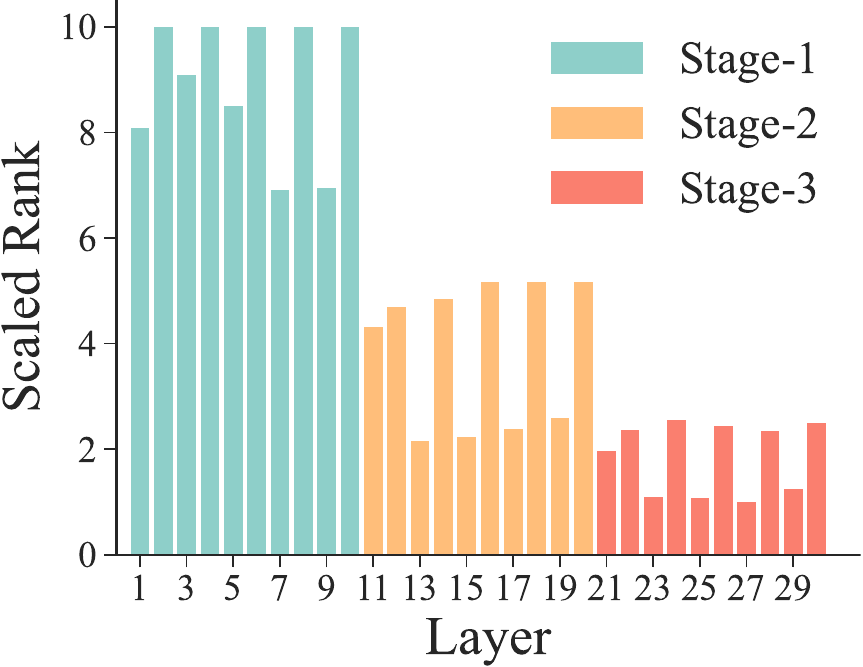}
				\label{fig:resnet32_rank}
			\end{minipage}
		}
		\subfloat[Entropy - ResNet-32]{
			\begin{minipage}[t]{0.3\textwidth}
				\centering
				\includegraphics[scale=0.3]{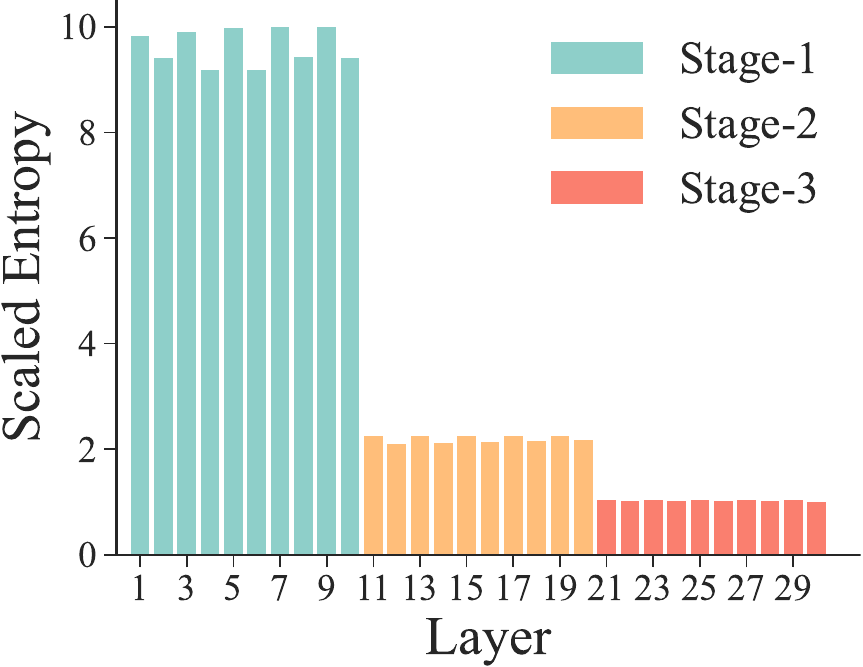}
				\label{fig:resnet32_entropy}
			\end{minipage}
		}
		\subfloat[Fusion Value - ResNet-32]{	
			\begin{minipage}[t]{0.3\textwidth}
				\centering
				\includegraphics[scale=0.3]{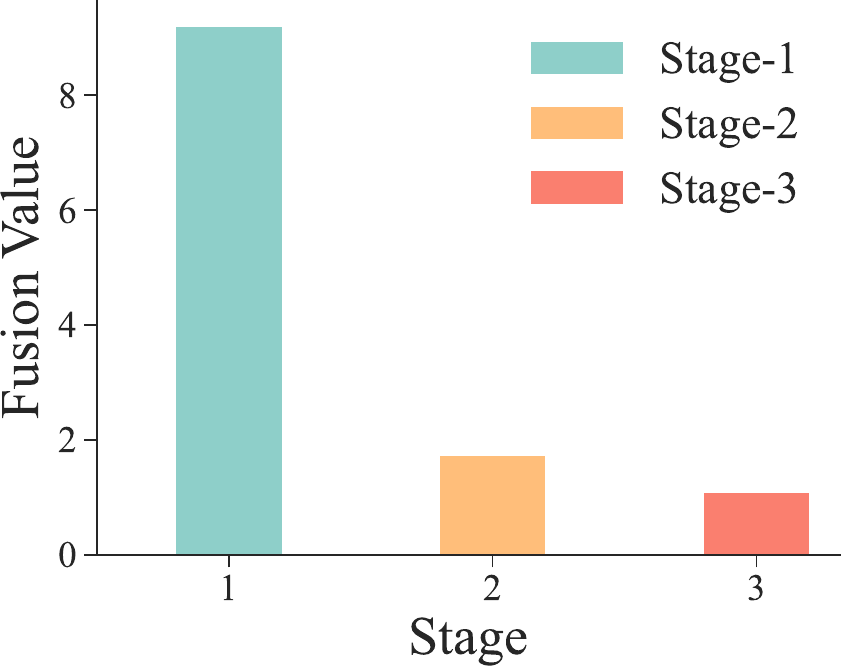}
				\label{fig:resnet32_fusion}
			\end{minipage}
		}
		
		
		\subfloat[Rank - ResNet-34]{
			\begin{minipage}[t]{0.3\textwidth}
				\centering
				\includegraphics[scale=0.3]{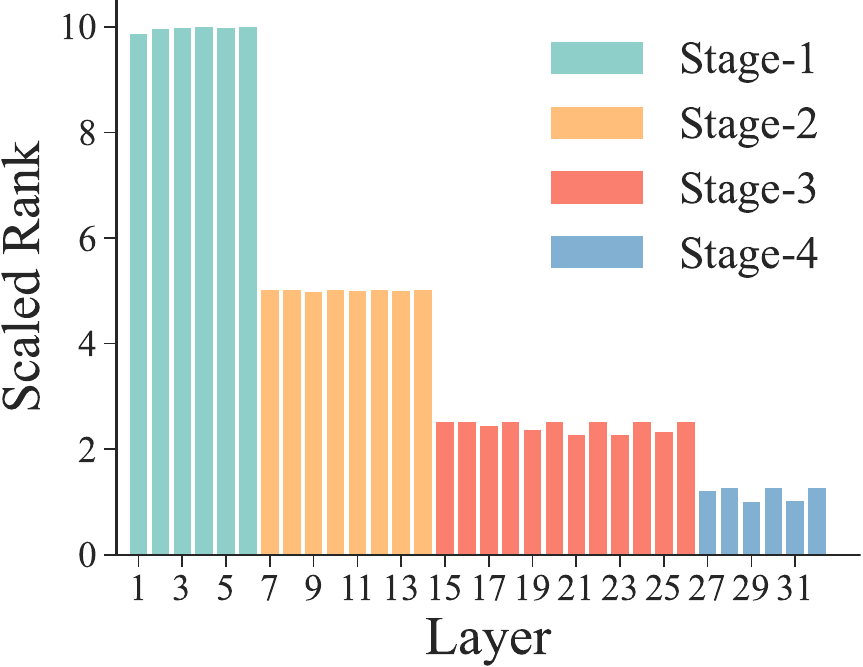}
				\label{fig:resnet34_rank}
			\end{minipage}
		}
		\subfloat[Entropy - ResNet-34]{
			\begin{minipage}[t]{0.3\textwidth}
				\centering
				\includegraphics[scale=0.3]{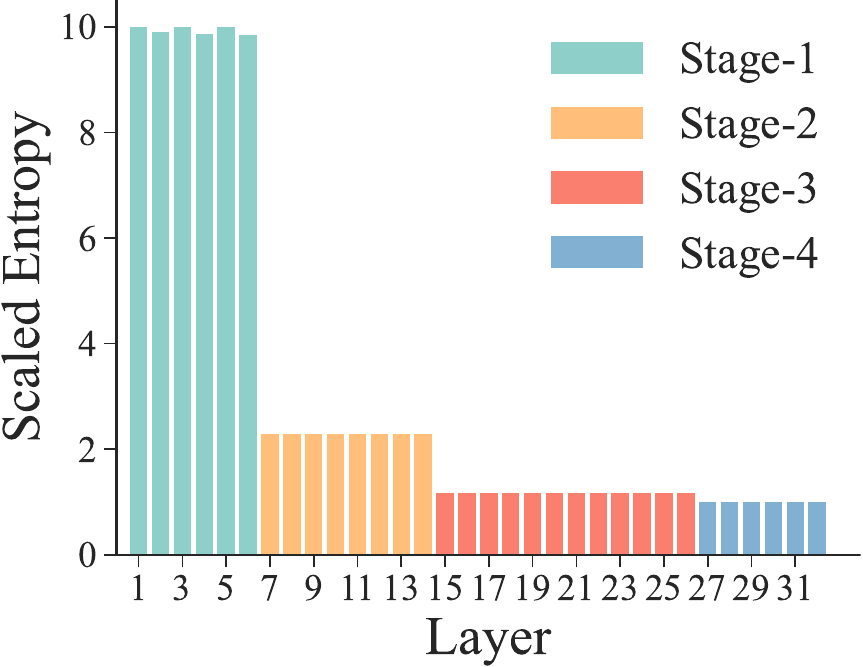}
				\label{fig:resnet34_entropy}
			\end{minipage}
		}
		\subfloat[Fusion Value - ResNet-34]{	
			\begin{minipage}[t]{0.3\textwidth}
				\centering
				\includegraphics[scale=0.3]{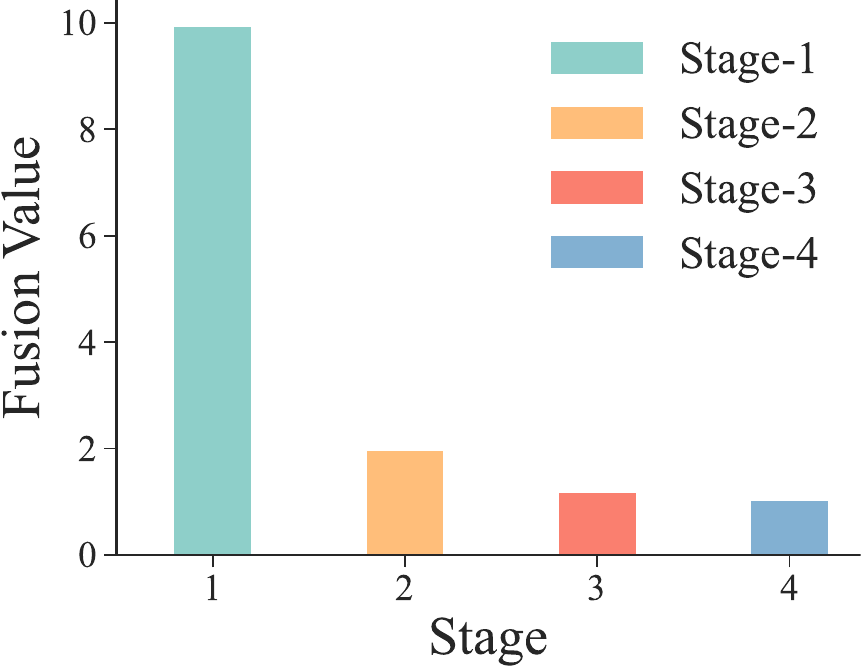}
				\label{fig:resnet34_fusion}
			\end{minipage}
		}
		
		\caption{Average statistics of rank and entropy for convolutional layer outputs and the corresponding fusion of them per layer inside a stage for ResNet with four depths.}
		\label{fig:rank_entropy_figure}
	\end{figure*}
	
	\subsection{Channel Pruning via Shapley Values}
	\label{subsec:sv} The concept of Shapley value comes from the field of cooperative game theory and is based on a scenario in which members cooperate in a team. They receive a reward that is meant to be equally allocated to each player based on their unique contributions, where a contribution is known as Shapley value \cite{shapley}. Shapley values are frequently used to interpret a ``black-box'' deep neural network \cite{captum, polynomial}, due to its rationality in evaluating the importance of features. Therefore, we extend the concept to the case of channel pruning: By considering a convolutional layer in a CNN-based model as a game in which the channels collaborate to produce an output, we can distribute the layer-wise outcomes to each of them. We suppose that a set $P = \{1, 2, \ldots, n\}$ is comprised of $n$ members who participate in cooperation, and the subset $s \subseteq P$ represents an alliance consisting of two or more members. If for any $s$, a corresponding real number $e(s)$ satisfies:
	\begin{equation}
		\begin{cases} e(\varnothing) = 0. \\
			\begin{split}
				\forall \text{ disjoint subsets } s_1, s_2 \subseteq P, \\ e(s_1 \cup s_2) \geq e(s_1) + e(s_2).
			\end{split}
		\end{cases}
	\end{equation} then $e(s)$ can be defined as a characteristic equation on $P$. The probability if the player $a$ will join the alliance $s\verb|\|\{a\}$ is calculated as:
	\begin{equation}
		\begin{split} v(\lvert s \rvert) & = \frac{(\lvert s \rvert - 1)!(n - \lvert s \rvert)!}{n!}
		\end{split}
	\end{equation} where $\lvert s \rvert$ represents the number of elements in $s$. The player $a$'s marginal contribution to all of the alliances which comprise $a$ is calculated as:
	\begin{equation} g_a(e) = \sum\limits_{s \in S_a} (e(s) - e(s \backslash \{a\})),
	\end{equation} where $S_a$ denotes the set that contains the player $a$ from all subsets, and $e(s) - e(s \backslash \{a\})$ denotes the marginal player's contribution, \ie, the player's contribution in alliance $s$. Thus, the Shapley value that the player $a$ generates is calculated as:
	\begin{equation}
		\label{eq:sv}
		\begin{split} w_a(e) & = \sum\limits_{s \in S_a} v(\lvert s \rvert) (e(s) - e(s \backslash \{a\}))	\propto g_a(e).
		\end{split}
	\end{equation} Since the Shapley value $w_a(e)$ is proportional to the marginal contribution $g_a(e)$, it reflects a player's contribution to the cooperation. In the case of convolutional neural networks, for simplicity, a deep CNN-based neural network $F$ can be written as:
	\begin{equation} F = F^{(1)} \circ F^{(2)} \circ \cdots F^{(L)}
	\end{equation} where $F^{(1)}, F^{(2)}, \cdots, F^{(L)}$ is the forward functions layer by layer. We regard $c_i$ channels $\{C_i^1, C_i^2, \ldots, C_i^{c_i}\}$ as $c_i$ members in the set $C_i$. The function $\hat{F}$ maps each subset $r \subseteq C_i$ of channels from activation outputs to real numbers for modeling the outcomes. The Shapley value of the channel $C_i^j$ is calculated as:
	\begin{equation}
		\begin{split} w_{C_i^j}(\hat{F}) = \sum \limits_{r \in S_{C_i^j}} \frac{(\lvert r \rvert - 1)!({c_i} - \lvert r \rvert)!}{{c_i}!} \\ (\hat{F}(r) - \hat{F}(r \backslash \{C_i^j\})). 
		\end{split}
	\end{equation} Therefore, $w_{C_i^j}(\hat{F})$ represents the importance of the channel, which provides us a reliable basis to discard the least important channels in network pruning. Thus, we can reformulate Eqn. (\ref{eq:opt}) as:
	\begin{equation}
		\begin{split}
			&\min\limits_{I_i^j} \sum\limits_{i=1}^{L} \sum\limits_{j=1}^{c_i} I_i^j w_{C_i^j}(\hat{F}),
			\\&
			s. t. \; \sum_{j=1}^{c_i} I_i^j = q_{i}.
		\end{split}
	\end{equation}
	
	\subsection{Pruning Schedules}
	\label{sec:pruning_schedules} In the pruning phase for image classification, three pruning schedules are employed by our method to prune the channels over the convolutional layers \cite{emerging}:
	\begin{enumerate}
		\item One-shot: We discard the channels from the convolutional layers with the lowest Shapley values representing contributions at once.
		\item Iterative static: We prune each layer and fine-tune the compact network for several epochs iteratively, and the unimportant channels are determined by the Shapley values calculated in the pre-trained model. ``Static'' implies that the channel contributions are represented by the initial Shapley values of them.
		\item Iterative dynamic: We prune and fine-tune the network similar to the iterative static paradigm, but the channel contributions are generated during the training and pruning procedure. ``Dynamic'' implies the Shapley values are calculated after fine-tuning, but not from the initial pre-trained network.
	\end{enumerate} Alg. \ref{alg:1} demonstrates our method with an iterative static pruning schedule, which we use in the experiments on CIFAR-10. First, we use the weights of the pre-trained model to initialize the network. Next, we propose a fusion function to get the overall indicator of rank and entropy for the convolutional layers under image batches. According to the fusion values, we assign the number of channels to be removed for the layers. By forward-passing the model to the batches of sampled images, we also obtain the Shapley values for the channels in the convolutional layers. During the pruning stage, we considered that the channels in a convolutional layer can be regarded as the team members in cooperation. Therefore, the Shapley values can be the reflection of their average marginal contributions to the layer output. After sorting the Shapley values, the lowest ones indicate the corresponding channels with the fewest contributions. Then, we discard the least important channels. After pruning a layer, to lessen the loss brought on by pruning, we fine-tune the model with several epochs. The procedure repeats until the final layer is pruned. Until the process of layer-wise pruning and fine-tuning ends, we retrain the pruned model to decrease the error caused by pruning. The accuracy of the compressed model is then restored. In our experiments, we demonstrated initializing with weights of pruned models outperformed random initialization, which we decided as the initialization method in re-training.
	
	\begin{algorithm}[ht!]
		\renewcommand{\algorithmicrequire}{\textbf{Input:}}
		\renewcommand{\algorithmicensure}{\textbf{Output:}}
		\caption{Pruning Schedule for Image Classification}
		\begin{algorithmic}[1]
			\REQUIRE pre-trained model $M$ with $L$ convolutional layers: $C = \{C_1, C_2, \ldots, C_L\}$, and training data: $D = \{D_1, D_2, \ldots, D_N\}$.
			
			\STATE \textbf{Initialize:} the weights of $M$.
			
			\STATE Feed sampled $B$ in $X$ to $M$ for inference to calculate: $O(C) = \{O(C_1), O(C_2), \ldots, O(C_L)\}$ and $w_{C}(\hat{F}) = \{w_{C_1}(\hat{F}), w_{C_2}(\hat{F}), \ldots, w_{C_L}(\hat{F})\}$.
			
			\STATE Assign the numbers of channels to be discarded $u = \{u_1, u_2, \ldots, u_L\}$ according to $O(C)$.

			\FOR{$i = 1; i \le L; i++$}
			
			\STATE{
				Sort the importance scores: $w_{C}(\hat{F}) = \{w_{C_i^1}(\hat{F}), w_{C_i^2}(\hat{F}), \ldots, w_{C_i^{c_i}}(\hat{F})\}$.
			}
			\STATE{
				Remove the last $u_i$ channels in $w_{C}(\hat{F})$.
			} 
			\STATE{
				Fine-tune the model with $D$.
			} 
			\ENDFOR
			
			\STATE Obtain the pruned and fine-tuned model $M^{'}$.
			
			\FOR{$epoch = 1; epoch \le epoch_{max}; epoch++$}
			
			\STATE{
				Retrain $M^{'}$ with $D$.
			} 
			\ENDFOR
			
			\ENSURE the compact model $M^{''}$.
		\end{algorithmic}
		\label{alg:1}
	\end{algorithm} The models in object detection contain not only ``backbone'' structures, but also ``neck'' and ``head''. As a result, the number of stages (blocks consisting of identical layers) becomes greatly larger than the ones in the image classification models, which results in many more hyper-parameters needing to be tuned. Semantic segmentation models incorporate not only 'base' structures to interpret input images but also 'upsampling' components to generate detailed classifications. This results in a greater depth of stages (blocks comprised of similar layers) compared to image classification models, necessitating a larger number of hyper-parameters to be finetuned. As is known to all, the more hyper-parameters we manually tune, the more errors it may cause. Finally, we decided to try another method in pruning models for those tasks that are more complex than image classification. Finally, we decided to try another method in pruning models for those tasks that are more complex than image classification. Inspired by SFP \cite{sfp} and Fisher \cite{fisher}, we employ the progressive pruning paradigm \cite{emerging} without pre-defining layer-wise pruning ratios, as summarized in Alg. \ref{alg:2}. Different from iterative pruning, we prune the network along with network training. In the training and pruning process, we calculate the Shapley values for all channels in the network and discard the least important one at every iteration. When the process ends, the unimportant channels are all removed, then we obtain a compact model. For the purpose of reconstructing the detection capability, we retrain the pruned model again.
	
	\begin{algorithm}[ht!]
		\renewcommand{\algorithmicrequire}{\textbf{Input:}}
		\renewcommand{\algorithmicensure}{\textbf{Output:}}
		\caption{Pruning Schedule for Object Detection and Semantic Segmentation}
		\begin{algorithmic}[1]
			\REQUIRE pre-trained model $M$, and training data: $D = \{D_1, D_2, \ldots, D_N\}$.
			
			\STATE \textbf{Initialize:} the weights of $M$.
			
			\REPEAT
			
			\STATE Sort the Shapley values of all channels in the network.
			
			\STATE Remove the least important channel and the corresponding filter.
			
			\UNTIL the target sparsity is achieved.
			
			\STATE Obtain the pruned and fine-tuned model $M^{'}$.
			
			\FOR{$epoch = 1; epoch \le epoch_{max}; epoch++$}
			
			\STATE{
				Fine-tune the model $M^{'}$ with $D$.
			} 
			\ENDFOR
			
			\ENSURE the compact model $M^{''}$.
		\end{algorithmic}
		\label{alg:2}
	\end{algorithm}
	
	\section{Experiments}
	
	\subsection{Experimental Settings}
	\label{subsec:settings}
	\textbf{Benchmark datasets and models:} We evaluate the performance of our pruning method on CIFAR-10 and ImageNet for image classification. The CIFAR-10 and ImageNet dataset contains 60,000 32 $\times$ 32 images with 10 classes and 1.28 million 224 $\times$ 224 images with 1,000 classes, respectively. We conduct the experiments on VGGNet with a plain structure, ResNet with a residual structure, and DenseNet with dense blocks. We randomly select 1,024 and 128 images for the architectures on CIFAR-10 and ImageNet for the purpose of estimating the average statistics of rank and entropy for convolutional layer outputs, and then we combine these two indicators to obtain the results. We set $[1, 10]$ as the scaling range of rank, entropy, and fusion values as Eqn. \ref{eq:fusion} introduced. In addition, we also evaluate the performance of our method in object detection on COCO2017, where the architectures are RetinaNet, FSAF, ATSS and PAA. COCO2017 contains 163,957 images with 80 categories, and 118,287 and 5,000 images are in the training set and validation set, respectively. Following the previous works \cite{ssd, faster-rcnn, yolo}, we train the models on the training set and evaluate their detection capabilities on the validation set.
	
	\textbf{Evaluation metrics:} We use Top-1 accuracy and Top-1/Top-5 accuracies on the test set to assess the image classification capabilities on CIFAR-10 and ImageNet, respectively. We use Float Points Operations (denoted as FLOPs) and the parameters to assess the computational overhead and the capacity of the pruned models, respectively. We evaluate the accuracy drop for performance, FLOPs reduction for acceleration and parameters reduction for compression in different methods. For object detection, we use mAP (mean average precision) to evaluate the detection capabilities of the models, and also use the memory footprint to estimate the scale of the models.
	
	\textbf{Configurations:} We use PyTorch \cite{pytorch} as the framework to implement our method. In the training process, we use SGD as the network optimizer and set 0.1 as the initial learning rate. Additionally, we set the batch size of 256. The momentum is 0.9 and weight decay is $10^{-4}$. On CIFAR-10, we train the model for 200 epochs, where the learning rate is divided by 10 at epochs 100 and 150. On ImageNet, we train the model for 90 epochs and the learning rate is divided by 10 at epochs 30 and 60. On COCO2017, we train the model for 12 epochs and the batch size is 2. The initial learning rate of 0.005. In terms of the pruning schedule, we adopt the iterative pruning manner on CIFAR-10, where the fine-tuning epochs are 20. Besides, we adopt the one-shot pruning schedule on ImageNet and progressive pruning paradigm on COCO2017. We use four NVIDIA RTX 3090 GPUs and four Tesla V100 GPUs for conducting the experiments.
	
	\subsection{Results and Analysis}
	
	\subsubsection{VGGNet, ResNet and DenseNet on CIFAR-10} The classification performance on CIFAR-10 is demonstrated in Tab. \ref{tab:results_cifar}.
	
	\begin{table*}[h!]
		\caption{Pruning results on CIFAR-10. ``Acc. Drop'' denotes the accuracy gap between the pruned and baseline model. To facilitate a more effective comparison of experimental results, we implemented the same method across two separate experiments. This approach is taken due to the considerable discrepancies observed in the reduction of FLOPs as reported in different papers.}
		\begin{center}
			
			\scalebox{1}{
				\setlength{\tabcolsep}{3.5mm}{
					\begin{tabular}{@{} c|c|cc|ccc@{}}
						\toprule
						Architecture& Algorithm & Baseline & Accuracy & Acc. Drop & FLOPs Drop & Params. Drop \\
						\midrule
						
						\multirow{5}{*}{VGG-16}
						& SSS \cite{sss} & 93.96\%  & 93.02\%  & 0.94\%  & 41.6\% & \textbf{73.8\%} \\
						& CP \cite{cp} & 93.26\%  & 90.80\%  & 2.46\%  & 50.6\% & -- \\
						& Ours & 93.91\%  & 93.17\%  & \textbf{0.74\%}  & \textbf{52.3\%} & 45.7\% \\ \cmidrule(l){2-7}
						& HRank \cite{hrank}               & 93.96\%           & 92.34\%           & 1.62\%           & \textbf{65.3\%}                & \textbf{82.1\%} \\
						& Ours & 93.91\%  & 93.38\%  & \textbf{0.53\%}  & 61.0\% & 50.7\% \\ \midrule
						
						\multirow{7}{*}{ResNet-20}  
						& SFP \cite{sfp}                 & 92.20\%           & 90.83\%           & 1.37\%           & 42.2\%           & --          \\
						& FPGM \cite{fpgm}                & 92.20\%           & 91.09\%           & 1.11\%           & 42.2\%           & --          \\
						& Ours & 91.73\%  & 91.38\%  & \textbf{0.35\%}  & \textbf{45.2\%}  & \textbf{40.3\%} \\
						\cmidrule(l){2-7} 
						& DSA \cite{dsa}                 & 92.17\%           & 91.38\%           & 1.06\%           & 50.3\%                & --          \\
						& FPGM \cite{fpgm}                & 92.20\%           & 90.44\%           & 1.76\%           & 54.0\%           & --          \\
						& PGMPF \cite{pgmpf}                & 92.89\%           & 91.54\%           & 1.35\%           & 54.0\%           & --          \\
						& Ours & 91.73\%  & 90.80\%  & \textbf{0.93\%}  & \textbf{57.9\%}  & \textbf{43.9\%} \\ \midrule
						
						\multirow{7}{*}{ResNet-32}
						& ACTD \cite{actd}                 & 93.18\%           & 93.27\%           & \textbf{-0.09\%}           & 38.0\%           & \textbf{49.0}          \\
						& SFP \cite{sfp}                 & 92.63\%           & 92.08\%           & 0.55\%           & 41.5\%           & --          \\
						& FPGM \cite{fpgm}                & 92.63\%           & 92.31\%           & 0.32\%           & 41.5\%           & --          \\
						& Ours & 92.63\%  & 92.64\%  & -0.01\%  & \textbf{46.8\%}  & 40.3\% \\ \cmidrule(l){2-7}
						& PScratch \cite{pscratch}                & 93.18\%           & 92.18\%           & 1.00\%          & 50.0\%                & --          \\
						& FPGM \cite{fpgm}                & 92.63\%           & 91.93           & \textbf{0.70\%}           & 53.2\%           & --          \\
						& Ours & 92.63\%  & 91.93\%  & \textbf{0.70\%}  & \textbf{58.0\%}  & \textbf{43.9\%} \\ \midrule
						
						\multirow{13}{*}{ResNet-56}
						& GAL \cite{gal}                 & 93.33\%           & 92.98\%           & 0.35\%          & 37.6\%           & 11.8\%          \\
						& ACTD \cite{actd}                 & 93.69\%           & 93.76\%           & -0.07\%          & 40.0\%           & \textbf{50.0}          \\
						& SFP \cite{sfp}                 & 93.59\%           & 93.78\%           & -0.19\%          & 41.1\%           & --          \\
						& GCNNA \cite{gcnna} & 93.72\%  & 93.85\%  & -0.13\% & 48.3\%  & 35.0\% \\
						& DECORE \cite{decore} & 93.26\%  & 93.26\%  & 0.00\% & 49.9\%  & 49.0\% \\
						& AMC \cite{amc} & 92.80\%  & 91.90\%  & 0.90\% & \textbf{50.0\%}  & -- \\
						& Ours & 93.39\%  & 93.60\%  & \textbf{-0.21\%} & 45.5\%  & 40.3\% \\
						\cmidrule(l){2-7} 
						& FPGM \cite{fpgm}                & 93.59\%           & 93.26\%           & 0.33\%          & 52.6\%                & --          \\
						& DBP \cite{dbp}                & 93.69\%           & 93.27\%           & 0.42\%          & 52.0\%                & 40.0\%          \\
						& DSA \cite{dsa}                 & 93.12\%           & 92.91\%           & 0.21\%           & 52.2\%                & --          \\
						& SFP \cite{sfp}                 & 93.59\%           & 93.35\%           & 0.24\%           & 52.6\%           & --          \\
						& Graph \cite{graph} & 93.27\%  & 93.38\%  & \textbf{-0.11\%} & \textbf{60.3\%}  & 43.0\% \\
						& Ours & 93.39\%  & 93.11\%  & 0.28\%  & 58.1\%  & \textbf{43.9\%} \\ \midrule
						
						\multirow{8}{*}{ResNet-110} 
						& SFP \cite{sfp}                 & 93.68\%           & 93.86\%           & -0.18\%          & 40.8\%          & --          \\
						& HRank \cite{hrank}               & 93.50\%           & 94.23\%           & -0.73\%          & 41.2\%          & 39.4\%          \\
						& Ours & 93.68\%  & 94.56\%  & \textbf{-0.88\%} & \textbf{45.6\%} & \textbf{40.4\%} \\ \cmidrule(l){2-7}
						& GAL \cite{gal}                & 93.50\%           & 92.74\%           & 0.76\%          & 48.5\%          & \textbf{44.8\%}          \\
						& FPGM \cite{fpgm}                & 93.68\%           & 93.74\%           & -0.16\%          & 52.3\%          & --          \\
						& HRank \cite{hrank}               & 93.50\%           & 93.36\%           & 0.14\%           & 58.2\%          & 59.2\%         \\
						& DECORE \cite{decore}               & 93.50\%           & 93.50\%           & 0.00\%           & \textbf{61.8\%}          & \textbf{64.8\%}         \\
						& Ours & 93.68\%  & 94.16\%  & \textbf{-0.48\%} & 58.1\% & 44.0\% \\ \midrule
						\multirow{4}{*}{DenseNet-40}
						& CC \cite{cc}               & 94.81\%          & 94.67\%           & \textbf{0.14\%}          & \textbf{47.0\%}         & 51.9\%          \\
						& Ours & 94.22\%  & 93.56\%  & 0.66\% & 44.4\% & \textbf{60.8\%} \\
						\cmidrule(l){2-7} 
						& HRank \cite{hrank}               & 94.81\%           & 93.53\%           & \textbf{1.28\%}           & 54.7\%          & 56.7\% \\
						& Ours & 94.22\%  & 92.54\%  & 1.68\% & \textbf{59.6\%} & \textbf{68.6\%} \\
						
						\bottomrule
						
					\end{tabular}
				}
			}
		\end{center}
		\label{tab:results_cifar}
	\end{table*}
	
	\textbf{VGGNet:} In comparison to SSS and CP, our method accelerates VGG-16 with a 52.3\% FLOPs reduction, achieving lower loss in accuracy (0.74\% \vs 0.94\% by SSS and 2.46\% by CP) and reducing larger FLOPs (52.3\% \vs 41.6\% by SSS and 50.6\% by CP). Additionally, our approach achieves a significantly lower accuracy loss (0.53\%) than HRank (1.62\%) while producing an acceleration ratio of 61.0\% and compression ratio of 50.7\%. Therefore, it shows that it is possible to exploit Shapley values as a criterion for discarding the unimportant channels.
	
	\textbf{ResNet:} For ResNet-20, our method is superior than SFP and FPGM \wrt loss in accuracy (0.35\% \vs 1.37\% by SFP and 1.11\% by FPGM) and FLOPs reduction (45.2\% \vs 42.2\% by SFP and 42.2\% by FPGM). Additionally, our method outperforms DSA, FPGM and PGMPF in accuracy drop (0.93\% \vs 1.06\% by DSA, 1.76\% by FPGM and 1.35\% by PGMPF), with larger FLOPs reduction (57.9\% \vs 50.3\% by DSA, 54.0\% by FPGM and 54.0\% by PGMPF). In comparison to SFP and FPGM, which obtain accuracy decreases of 0.55\% and 0.32\%, respectively, our method increases the accuracy by 0.01\% for ResNet-32. Moreover, our method reduces more FLOPs (38.0\% by ACTD, 46.8\% \vs 41.5\% by SFP and 41.5\% by FPGM). Compared with PScratch and FPGM, our method yields a loss in accuracy of 0.70\% lower than PScratch (1.00\%) and equal to FPGM (0.70\%). In addition, 43.9\% of parameters and 58.0\% of FLOPs are reduced by our method, where the FLOPs reductions are larger than Pscratch (50.0\%) and FPGM (53.2\%). For pruning ResNet-56, our method improves the accuracy more than ACTD, SFP and GCNNA (0.21\% \vs 0.07\% by ACTD, 0.19\% by SFP and 0.13\% by GCNNA), while GAL and AMC obtain the loss in accuracy of 0.35\% and 0.90\%, respectively, and DECORE retains the accuracy. In addition, under a larger acceleration ratio of 58.1\% and compression ratio of 43.9\%, our method obtains a drop in accuracy less than FPGM and DBP (0.28\% \vs 0.33\% by FPGM and 0.42\% by DBP), while Graph increases the accuracy by 0.11\% with parameters reduction of 43.0\%. Compared to the methods pruning ResNet-110, our method performs the best in accuracy. Specifically, our method increases the accuracy of 0.88\%, outperforming SFP (0.18\%) and HRank (0.73\%), reducing more FLOPs (45.6\% \vs 41.2\% by HRank and 40.8\% by SFP). Additionally, our method increases the accuracy larger than FPGM (0.48\% \vs 0.16\% by FPGM) with greater FLOPs reduction (58.1\% \vs 52.3\%), while GAL degrades the accuracy by 0.76\% and removes fewer FLOPs (48.5\%), and DECORE retains the accuracy with 61.8\% FLOPs drop higher than our method. HRank obtains a slightly higher FLOPs reduction of 58.2\% than our method, but it harms the accuracy of 0.14\%.
	
	\textbf{DenseNet:} Based on the results for pruning DenseNet-40, we observe that our method has the potential to remove more parameters. To be specific, although CC obtains an accuracy loss of 0.14\%, it achieves a FLOPs reduction that is close to our method (44.4\% \vs 47.0\% by CC) but a compression ratio which is lower than our method (60.8\% \vs 51.9\% by CC). In addition, our method degrades the accuracy drop slightly more than HRank (1.68\% \vs 1.28\% by HRank), but outperforms HRank \wrt FLOPs reduction (59.6\% \vs 54.7\% by HRank) and parameters reduction (68.6\% \vs 53.8\% by HRank). Thus, our method can more effectively compress models with dense blocks.
	
	\subsubsection{ResNet on ImageNet} The classification performance on ImageNet is demonstrated in Tab. \ref{tab:results_imagenet}.
	
	\begin{table*}[h!]
		\centering
		\caption{Pruning results on ImageNet. Each column \wrt accuracy includes Top-1/Top-5 accuracies. To facilitate a more effective comparison of experimental results, we implemented the same method across two separate experiments. This approach is taken due to the considerable discrepancies observed in the reduction of FLOPs as reported in different papers.}
		\label{tab:results_imagenet}
		
		\begin{center}
			
			\scalebox{1}{
				\setlength{\tabcolsep}{2.3mm}{
					\begin{tabular}{@{} c|c|cc|ccc@{}}
						\toprule
						Architecture& Algorithm & Baseline & Accuracy & Acc. Drop & FLOPs Drop & Params. Drop \\
						\midrule
						
						\multirow{4}{*}{ResNet-18}  
						& SFP \cite{sfp}            & 70.28\%/89.63\%                 & 67.10\%/87.78\%                 & 3.18\%/1.85\%             & \textbf{41.8\%}               & --          \\
						& FPGM \cite{fpgm}          & 70.28\%/89.63\%                 & 68.34\%/88.53\%                 & 1.94\%/1.10\%             & \textbf{41.8\%}              & --          \\
						& Ours & 69.76\%/89.08\%        & 68.27\%/88.26\%        & \textbf{1.49\%}/\textbf{0.82\%}    & 40.7\% & \textbf{40.6\%} \\  \cmidrule(l){2-7} 
						& Ours & 69.76\%/88.10\%        & 67.96\%/89.08\%        & \textbf{1.80\%}/\textbf{0.98\%}    & \textbf{45.1\%} & \textbf{41.7\%} \\ \midrule
						
						\multirow{5}{*}{ResNet-34} & SFP \cite{sfp}            & 73.92\%/91.62\%                 & 71.83\%/90.33\%                 & 2.09\%/1.29\%             & 41.1\%               & --          \\
						& FPGM \cite{fpgm}          & 73.92\%/91.62\%                 & 72.54\%/91.13\%                 & 1.38\%/\textbf{0.49\%}             & 41.1\%               & --          \\
						& CHEX \cite{chex}          & 73.92\%/91.62\%                 & 72.7\%/-                 & 1.22\%/-        & {\bf 45.9\%}               & --          \\
						& Ours & 73.31\%/91.42\%        & 72.74\%/90.86\%        & \textbf{0.57\%}/0.56\%    &  \textbf{45.2\%} & \textbf{36.3\%} \\  \cmidrule(l){2-7} 
						& PGMPF \cite{pgmpf} & 73.27\%/91.43\%        & 71.59\%/90.45\%        & 1.68\%/0.98\%    & \textbf{52.7\%} & -- \\
						& Ours & 73.31\%/91.42\%        & 72.25\%/90.63\%        & \textbf{1.06\%}/\textbf{0.73\%}    & 50.1\% & \textbf{36.0\%} \\ \midrule
						\multirow{13}{*}{ResNet-50}
						& DSA \cite{dsa}            & 76.02\%/92.86\%                 & 75.10\%/92.45\%                 & 0.92\%/0.41\%             & 40.0\%               & --          \\
						& SFP \cite{sfp}            & 76.15\%/92.87\%                 & 74.61\%/92.06\%                 & 1.54\%/0.81\%             & 41.8\%               & --          \\
						& FPGM \cite{fpgm}          & 76.15\%/92.87\%                 & 75.59\%/92.63\%                 & 0.56\%/0.24\%             & 42.2\%               & --          \\
						& GAL \cite{gal}            & 76.15\%/92.87\%                 & 71.95\%/90.94\%                 & 4.20\%/1.93\%             & 43.0\%          & 16.9\%          \\
						& HRank \cite{hrank}        & 76.15\%/92.87\%                 & 74.98\%/92.33\%                 & 1.17\%/0.54\%             & 43.7\%          & 36.7\% \\ 
						& DECORE \cite{decore}            & 76.15\%/92.87\%                 & 74.58\%/92.18\%                 & 1.57\%/0.69\%             & \textbf{44.7\%}               & \textbf{42.3\%}          \\
						& Ours & 76.13\%/92.86\%             & 75.70\%/92.75\%             & \textbf{0.43\%}/\textbf{0.11\%}        & 41.6\% & 35.0\% \\
						\cmidrule(l){2-7} 
						& DSA \cite{dsa}            & 76.02\%/92.86\%                 & 74.69\%/92.06\%                 & 1.33\%/0.80\%             & 50.0\%               & --          \\
						& TPP \cite{tpp}            & 76.13\%/--                 & 75.60\%/--                 & 0.53\%/--             & --               & --          \\
						& Fisher \cite{fisher} & 76.79\%/--        & 76.42\%/--       & \textbf{0.37\%}/--    & 50.0\% & -- \\
						& GNN-RL \cite{gnnrl} & 76.10\%/--        & 74.28\%/--       & 1.82\%/--    & 53.0\% & -- \\
						& PGMPF \cite{pgmpf} & 76.01\%/92.93\%        & 75.11\%/92.41\%       & 0.90\%/0.52\%    & \textbf{53.5\%} & -- \\
						& Ours & 76.13\%/92.86\%        & 75.29\%/92.47\%        & 0.84\%/\textbf{0.39\%}    & 50.4\% & \textbf{44.2\%} \\
						\midrule
						
						\multirow{4}{*}{ResNet-101} 
						& FPGM \cite{fpgm}           & 77.37\%/93.56\%                 & 77.32\%/93.56\%                 & 0.05\%/0.00\%             & 42.2\%               & --          \\
						& Ours & 77.37\%/93.55\%        & 77.35\%/93.59\%        & \textbf{0.02\%}/\textbf{-0.04\%}   & \textbf{43.7\%} & \textbf{42.6\%} \\ \cmidrule(l){2-7} 
						& Rethinking \cite{rethinking} & 77.37\%/--        & 75.27\%/--        & 2.10\%/--   & 47.0\% & -- \\
						& Ours & 77.37\%/93.55\%        & 76.10\%/92.94\%        & \textbf{1.27\%}/\textbf{0.61\%}   & \textbf{54.4\%} & \textbf{54.0\%} \\
						
						\bottomrule
				\end{tabular}}
			}
		\end{center}
	\end{table*} For ResNet-18, our method yields Top-1/Top-5 accuracy drops of 1.49\%/0.82\%, outperforming SFP (3.18\%/1.85\%) and FPGM (1.94\%/1.10\%). Moreover, our method achieves a FLOPs reduction of 40.7\% and a parameters reduction of 40.6\%. In addition, our method still outperforms SFP and FPGM with greater FLOPs and parameter reductions of 45.1\% and 41.7\%, respectively, with Top-1/Top-5 accuracy drops of 1.80\%/0.98\%. For ResNet-34, our method obtains a Top-1 accuracy drop of 0.57\%, which is better than SFP (2.09\%) and FPGM (1.38\%), removing larger FLOPs reduction (45.2\% \vs 41.1\% by SFP and 41.1\% by FPGM) and reducing 36.3\% parameters. Additionally, our method achieves Top-1/Top-5 accuracy drops of 1.06\%/0.73\% with a FLOPs reduction of 50.1\% and a parameters drop of 36.0\%, outperforming the Top-1/Top-5 accuracy drops of PGMPF (1.68\%/0.98\%) with 52.7\% FLOPs reduction. ResNet-50 is a popular network for analyzing the performance of pruning methods. Therefore, more methods for comparison are listed. Our method obtains  Top-1/Top-5 accuracy drops of 0.43\%/0.11\%, outperforming DSA (0.92\%/0.41\%), SFP (1.54\%/0.81\%), FPGM (0.56\%/0.24\%), GAL (4.20\%/1.93\%), HRank (1.17\%/0.54\%) and DECORE (1.57\%/0.69\%), among which the FLOPs reductions are similar. In addition, our method achieves Top-1/Top-5 accuracy drops of 0.84\%/0.39\%, outperforming DSA (1.33\%/0.80\%) with a parameter reduction of 44.2\%. In addition, our method achieves a loss in Top-1 accuracy slightly higher than TPP (0.53\%) and Fisher (0.37\%), lower than GNN-RL (1.82\%) and PGMPF (0.90\%), but with a larger FLOPs reduction of 50.4\% than DSA (50.0\%) and Fisher (50.0\%), and a parameters reduction of 44.2\%. ResNet-101 is a network with a large quantity of parameters, and our method outperforms FPGM in Top-1/Top-5 accuracy drops (0.02\%/-0.04\% \vs 0.05\%/0.00\% by FPGM) but removing more FLOPs (43.7\% \vs 42.2\% by FPGM). Additionally, when reducing 54.4\% FLOPs and removing 54.0\% parameters, our method yields Top-1/Top-5 accuracy drops of 1.27\%/0.61\%. On the contrary, Rethinking harms the Top-1 accuracy by 2.10\% and reduces fewer FLOPs (47.0\%).
	
	\subsubsection{RetinaNet, FSAF, ATSS and PAA on COCO2017} The detection performance on COCO2017 is demonstrated in Tab. \ref{tab:results_coco} comparing our method with Fisher \cite{fisher} of 50\% MACs ($1MACs = 2FLOPs$) drops under the same baseline \cite{mmdetection}. Then we compare mAP, amount of parameters and memory footprint of them.
	
	\begin{table*}[h!]
		\centering
		\caption{Pruning results on COCO2017. The model name without suffix denotes the unpruned model. ``-P'' and ``-F'' indicate the pruned model achieved by the methods which normalize importance scores by parameters reduction and FLOPs reduction, respectively. ``-N'' indicates the method employs only the scoring metrics without normalization. }
		\label{tab:results_coco}
		
		\begin{center}
			\scalebox{1}{
				\setlength{\tabcolsep}{2.3mm}{
					\begin{tabular}{@{} c|l|ccc|c@{}}
						\toprule
						Architecture & Algorithm & mAP & Params. & Memory& MACs  \\ \midrule
						
						\multirow{7}{*}{RetinaNet}
						& Baseline & 36.5\%  & 37.96M  & 297.40M & 238.50G   \\ \cmidrule(l){2-6}
						& Fisher - P & 36.5\%& \textbf{26.34M}  & \textbf{190.20M}  & 119.25G   \\
						& Ours - P & \textbf{37.3\%} & 26.27M  & 201.18M  & 119.25G   \\
						\cmidrule(l){2-6}
						& Fisher - F & 36.8\%& 26.18M  & 254.10M  & 119.25G   \\
						& Ours - F & \textbf{37.6\%} & \textbf{25.55M}   & \textbf{195.65M} & 119.25G \\
						\cmidrule(l){2-6}
						& Fisher - N & \textbf{35.8\%}  & 14.18M  & 244.20M & 119.25G   \\
						& Ours - N & 35.6\%  & \textbf{13.49M}  & \textbf{103.59M} & 119.25G  \\
						\midrule
						
						\multirow{7}{*}{FSAF}
						& Baseline & 37.4\%  & 36.41M  & 283.10M & 205.50G  \\ \cmidrule(l){2-6}
						& Fisher - P & 37.3\%  & \textbf{24.09M}  & \textbf{171.30M} & 102.75G  \\
						& Ours - P & \textbf{37.6\%}  & 24.12M  & 184.73M & 102.75G  \\
						\cmidrule(l){2-6}
						
						& Fisher - F & 37.6\%  & 23.19M  & 233.60M & 102.75G \\
						& Ours - F & \textbf{38.0\%}  & \textbf{22.30M}  & \textbf{170.84M} & 102.75G \\
						\cmidrule(l){2-6}
						
						& Fisher - N & \textbf{36.4\%}  & 19.23M  & 242.20M & 102.75G \\
						& Ours - N & 35.9\%  & \textbf{12.24M}  & \textbf{94.06M} & 102.75G \\
						\midrule
						
						\multirow{7}{*}{ATSS}
						& Baseline & 38.1\%  & 32.29M  & 283.10M & 204.40G \\ \cmidrule(l){2-6}
						& Fisher - P & 38.0\%  & \textbf{22.20M}  & 177.10M & 102.20G \\
						& Ours - P & \textbf{38.6\%}  & 22.23M  & \textbf{170.33M} & 102.20G \\
						\cmidrule(l){2-6}
						
						& Fisher - F & 38.3\%  & 21.54M  & 236.90M & 102.20G \\
						& Ours - F & \textbf{39.1\%}  & \textbf{20.49M}  & \textbf{157.05M} & 102.20G \\
						\cmidrule(l){2-6}
						
						& Fisher - N & \textbf{36.7\%}  & 12.41M  & 228.50M & 102.10G  \\
						& Ours - N & 36.5\%  & \textbf{11.55M}  & \textbf{88.83M} & 102.20G \\
						\midrule
						
						\multirow{7}{*}{PAA}
						& Baseline & 39.0\%  & 32.29M  & 283.10M & 204.40G \\ \cmidrule(l){2-6}
						& Fisher - P & 39.4\%  & \textbf{23.00M}  & \textbf{174.90M} & 102.10G  \\
						& Ours - P & \textbf{39.7\%}  & 23.06M  & 176.66M & 102.20G \\
						\cmidrule(l){2-6}
						
						& Fisher - F & 39.6\%  & 21.95M  & 235.50M & 102.10G \\
						& Ours - F & \textbf{40.0\%}  & \textbf{20.86M}  & \textbf{159.91M} & 102.20G \\
						\cmidrule(l){2-6}
						
						& Fisher - N & \textbf{38.5\%}  & 13.19M  & 228.40M & 102.20G \\
						& Ours - N & 37.7\%  & \textbf{11.86M}  & \textbf{91.14M} & 102.20G \\
						
						\bottomrule
				\end{tabular}}
			}
		\end{center}
	\end{table*}
	
	\textbf{RetinaNet:} For RetinaNet, when normalizing the Shapley values with run-time parameters reduction, our method achieves an mAP of 37.3\%, higher than the baseline model (36.5\%), while Fisher obtains no increase in mAP (36.5\%). In addition, our method yields a higher mAP (37.6\% \vs 36.8\%), fewer parameters (25.55M \vs 26.18M) and less memory footprint (195.65M \vs 254.10M) under the normalization of FLOPs reduction. Besides, without normalization of the importance scores, our method obtains an mAP similar to Fisher (35.6\% \vs 35.8\%), but fewer remaining parameters (13.49M \vs 14.18M) and notably less memory footprint (103.59M \vs 244.20M).
	
	\textbf{FSAF:} For FSAF, when normalizing the Shapley values with run-time parameters reduction, our method achieves an mAP of 37.6\%, higher than the baseline model (37.4\%), while Fisher harms the mAP of 37.3\%. Besides, our method yields a higher mAP (38.0\% \vs 37.6\%), fewer parameters (22.30M \vs 23.19M) and less memory footprint (170.84M \vs 233.60M) under the normalization of FLOPs reduction. Besides, without normalization of the importance scores, our method obtains an mAP of 35.9\%, lower than Fisher (35.6\%), but fewer remaining parameters (12.24M \vs 19.23M) and remarkably less memory footprint (94.08M \vs 242.20M).
	
	\textbf{ATSS:} For ATSS, when normalizing the Shapley values with run-time parameters reduction, our method achieves an mAP of 38.6\%, higher than the baseline model (38.1\%), while Fisher slightly degrades the mAP of 38.0\%. Besides, our method yields a higher mAP (39.1\% \vs 38.3\%), fewer parameters (20.49M \vs 23.19M) and less memory footprint (157.05M \vs 236.90M) under the normalization of FLOPs reduction. Besides, without normalization of the importance scores, our method obtains an mAP of 36.5\%, lower than Fisher (36.7\%), but fewer remaining parameters (11.55M \vs 19.23M) and remarkably less memory footprint (88.83M \vs 228.50M).
	
	\textbf{PAA:} For PAA, when normalizing the Shapley values with run-time parameters reduction, our method obtains higher mAP than Fisher (39.7\% \vs 39.4\%), where the performance of the two methods both outperform the baseline model (39.0\%). In addition, our method yields a higher mAP (40.0\% \vs 39.6\%), fewer parameters (20.86M \vs 21.95M) and less memory footprint (159.91M \vs 235.50M) under the normalization of FLOPs reduction. Besides, without normalization of the importance scores, our method obtains an mAP of 37.7\%, lower than Fisher (38.5\%), but fewer remaining parameters (11.85M \vs 13.19M) and remarkably less memory footprint (91.14M \vs 228.40M).
	
	\begin{table*}[h!]
		
		\captionsetup{labelfont={color=red}}
		\caption{Pruning transformers on CIFAR-10/100. ``Acc. Drop'' denotes the accuracy gap between the pruned and baseline model.}
		\begin{center}
			\scalebox{1}{
				\setlength{\tabcolsep}{3.5mm}{
					
					\begin{tabular}{@{} c|c|cc|ccc@{}}
						\toprule
						Dataset& Architecture & Baseline & Accuracy & Acc. Drop & FLOPs Drop & Params. Drop \\
						\midrule
						
						\multirow{5}{*}{CIFAR-10}
						& DeiT-T & 98.01\%  & 96.13\%  & 1.88\%  & 47.8\% & 47.7\% \\
						& DeiT-S & 98.87\%  & 97.75\%  & 1.12\%  & 47.8\% & 47.7\% \\
						& DeiT-B & 99.10\%  & 98.61\%  & 0.49\%  & 47.8\% & 47.7\% \\ 
						& ViT-B/16 & 99.03\%  & 97.52\%  & 1.51\%  & 51.1\% & 51.4\% \\
						& ViT-B/32 & 98.67\%  & 96.89\%  & 1.78\%  & 51.1\% & 51.4\% \\
						\midrule
						
						\multirow{4}{*}{CIFAR-100}
						& DeiT-T & 87.37\%  & 85.24\%  & 2.13\%  & 47.8\% & 47.7\% \\
						& DeiT-B & 91.49\%  & 89.68\%  & 1.81\%  & 47.8\% & 47.7\% \\ 
						& ViT-B/16 & 92.93\%  & 89.47\%  & 3.46\%  & 51.1\% & 51.4\% \\
						& ViT-B/32 & 92.12\%  & 88.96\%  & 3.16\%  & 51.1\% & 51.4\% \\
						\midrule
						
						\multirow{1}{*}{MNLI}
						& BERT$_{base}$ & 84.31\%  & 82.11\%  & 2.20\%  & 48.1\% & 42.2\% \\
						
						\bottomrule
						
					\end{tabular}
				}
			}
		\end{center}
		\label{tab:transformers_cifar}
	\end{table*}
	
	{\textit{4) DeiT\cite{deit} and ViT\cite{vit} on CIFAR-10\cite{cifar}, CIFAR-100\cite{cifar}. BERT$_{base}$\cite{bert} on MNLI\cite{mnli}:} The classification performance on
		ImageNet is demonstrated in Tab. \ref{tab:transformers_cifar}. We applied our pruning method to various Transformer architectures, including DeiT (Tiny, Small, Base), ViT (B/16, B/32), and the BERT base model. The experiments are conducted on several datasets, including CIFAR-10, CIFAR-100, and MNLI. DeiT: The DeiT models show a robust performance on both CIFAR-10 and CIFAR-100 datasets after pruning through our method. On CIFAR-10, the DeiT-T, DeiT-S, and DeiT-B models experience a relatively small accuracy drop of 1.88\%, 1.12\%, and 0.49\% respectively, compared to their baseline models. This is an impressive result considering the significant reduction in FLOPs and parameters by 47.8\% and 47.7\%, respectively, suggesting a good balance between efficiency and performance. The CIFAR-100 dataset follows a similar trend, with a slightly higher accuracy drop of 2.13\% for DeiT-T and 1.81\% for DeiT-B, while maintaining the same level of FLOPs and parameters reduction. ViT: The pruned ViT models also display impressive performance on both datasets. On CIFAR-10, the ViT-B/16 and ViT-B/32 models see an acceptable accuracy drop of 1.51\% and 1.78\% respectively, while achieving higher FLOPs and parameter reductions of 51.1\% and 51.4\% compared to DeiT models. However, on the CIFAR-100 dataset, the models experience a slightly more pronounced accuracy drop (3.46\% and 3.16\% respectively), which may suggest a trade-off between complexity reduction and performance in more challenging tasks. BERT: When pruning is applied to the BERT$_{base}$ model on the MNLI dataset, the result is a modest accuracy drop of 2.2\%. However, this comes with a significant reduction in computational complexity, given the FLOPs reduction of 48.1\%, and a substantial decrease in model size, as signified by the 42.2\% decrease in parameters. This suggests that even for complex language tasks, our pruning strategy can effectively reduce the model's resource demands while maintaining a competitive performance level.}
	
	\begin{table*}[ht!]
		
		\captionsetup{labelfont={color=red}}
		
		\caption{Pruning UNets in semantic segmention. ``Acc. Drop'' denotes the accuracy gap between the pruned and baseline model.}
		\begin{center}
			\scalebox{1}{
				\setlength{\tabcolsep}{3.5mm}{
					\begin{tabular}{@{} c|c|cc|ccc@{}}
						\toprule
						Dataset& Architecture & Baseline & Accuracy & Acc. Drop & FLOPs Drop & Params. Drop \\
						\midrule
						
						\multirow{2}{*}{BUSI\_2}
						& UNet & 64.55\%  & 64.17\%  & 0.38\%  & 49.0\% & 50.6\% \\
						& UNet++ & 63.45\%  & 61.01\%  & 2.44\%  & 49.2\% & 50.7\% \\
						\midrule
						
						\multirow{2}{*}{BUSI\_3}
						& UNet & 60.61\%  & 59.29\%  & 1.32\%  & 49.0\% & 50.6\% \\
						& UNet++ & 59.06\%  & 60.27\%  & -1.21\%  & 49.2\% & 50.7\% \\
						\midrule
						
						\multirow{2}{*}{DSB}
						& UNet & 84.25\%  & 85.34\%  & -1.07\%  & 49.0\% & 50.6\% \\
						& UNet++ & 85.03\%  & 84.88\%  & 1.15\%  & 49.2\% & 50.7\% \\
						\midrule
						
						\multirow{2}{*}{ISIC}
						& UNet & 80.48\%  & 80.41\%  & 0.07\%  & 49.0\% & 50.6\% \\
						& UNet++ & 79.63\%  & 79.88\%  & -0.25\%  & 49.2\% & 50.7\% \\
						
						\bottomrule
						
					\end{tabular}
				}
			}
		\end{center}
		\label{tab:segmentation}
	\end{table*}
	
	{	{\textit{5) UNet\cite{unet} and UNet++\cite{unet++} on BUSI\_2 \cite{busi}, BUSI\_3 \cite{busi}, DSB (data science bowl) and ISIC\cite{isic}}: For these four datasets, we observed that the drop in accuracy of the pruned models is very small. In some cases, the accuracy is even improved as shown in Tab. \ref{tab:segmentation}. 
			
	BUSI\_2: For the BUSI\_2 dataset, when pruning UNets, our method with UNet architecture results in a slight accuracy drop of 0.38\% (64.55\% to 64.17\%), while achieving a FLOPs reduction of 49.0\% and parameters reduction of 50.6\%. For UNet++, the accuracy drops by a larger margin of 2.44\% (63.45\% to 61.01\%), but also with a FLOPs reduction of 49.2\% and parameters reduction of 50.7\%. BUSI\_3: For the BUSI\_3 dataset, our method with UNet architecture results in an accuracy drop of 1.32\% (60.61\% to 59.29\%), while achieving a FLOPs reduction of 49.0\% and parameters reduction of 50.6\%. Interestingly, UNet++ shows a slight increase in accuracy by 1.21\% (59.06\% to 60.27\%), with a FLOPs reduction of 49.2\% and parameters reduction of 50.7\%. DSB: For the DSB dataset, our method with UNet architecture results in an increase in accuracy by 1.07\% (84.25\% to 85.34\%), while achieving a FLOPs reduction of 49.0\% and parameters reduction of 50.6\%. For UNet++, the accuracy slightly decreases by 1.15\% (85.03\% to 84.88\%), but also with a FLOPs reduction of 49.2\% and parameters reduction of 50.7\%. ISIC: For the ISIC dataset, our method with UNet architecture results in a negligible accuracy drop of 0.07\% (80.48\% to 80.41\%), while achieving a FLOPs reduction of 49.0\% and parameters reduction of 50.6\%. For UNet++, the accuracy slightly increases by 0.25\% (79.63\% to 79.88\%), with a FLOPs reduction of 49.2\% and parameters reduction of 50.7\%.	}}
	
	\section{Ablation Studies}
	
	\subsection{Performance w.r.t Pruning Schedules} Tab. \ref{tab:diff_prune_schedules} shows the performance of the pruned VGGNet, ResNet and DenseNet with three pruning schedules on CIFAR-10. The one-shot pruning schedule saves the greatest time because it prunes the network all at once. however, in our experiments, only VGG-16 with the one-shot schedule under the FLOPs reduction of 52.3\% and parameters reduction of 45.7\% outperforms the other two schedules. The accuracy of the pruned model is not substantially harmed by the iterative static and iterative dynamic pruning schedules, even increases on ResNet-32 with 46.8\% FLOPs reduction, ResNet-56 with 45.5\% FLOPs reduction and ResNet-110 with 45.6\% and 58.1\% FLOPs reductions. Nevertheless, it should be noted that the iterative static pruning schedule attains a best accuracy-efficiency trade-off since no calculation overhead of the Shapley values on each convolutional layer is required during the process of layer-wise fine-tuning the compact network.
	
	\begin{table*}[ht!]
		\caption{Pruning results \wrt pruning schedules on CIFAR-10. ``OS'', ``IS'' and ``ID'' denote one-shot, iterative static and iterative dynamic pruning schedules introduced in Sec. \ref{sec:pruning_schedules}, respectively.}
		\begin{center}
			\scalebox{1}{
				\setlength{\tabcolsep}{3.5mm}{
					\begin{tabular}{@{} c|c|cc|c|cc@{}}
						\toprule
						Architecture& Algorithm & Baseline & Accuracy & Acc. Drop & FLOPs Drop & Params. Drop \\
						\midrule
						
						\multirow{6}{*}{VGG-16}     & OS               & \multirow{3}{*}{93.91\%}           & 93.27\%           & \textbf{0.64\%}  & \multirow{3}{*}{52.3\%} & \multirow{3}{*}{45.7\%} \\
						& IS               &           & 93.17\%         & 0.74\%           &                       &                       \\
						& ID               &           & 93.10\%           & 0.81\%           &                       &                       \\ \cmidrule(l){2-7} 
						& OS               & \multirow{3}{*}{93.91\%}      & 92.92\% & 0.99\%           & \multirow{3}{*}{61.0\%} & \multirow{3}{*}{50.7\%} \\
						& IS               &           & 93.24\%           & 0.67\%           &                       &                       \\
						& ID               &           & 93.41\%           & \textbf{0.50\%}  &                       &                       \\ \midrule
						\multirow{6}{*}{ResNet-20}  & OS               & \multirow{3}{*}{91.73\%}            & 91.05\%           & 0.68\%           & \multirow{3}{*}{45.2\%} & 
						\multirow{3}{*}{40.3\%} \\
						& IS               &            & 91.38\%           & \textbf{0.35\%}  &                       &                       \\
						& ID               &            & 91.14\%           & 0.59\%           &                       &                       \\ \cmidrule(l){2-7} 
						& OS               &   
						\multirow{3}{*}{91.73\%}         & 90.70\%           & 1.03\%  & \multirow{3}{*}{57.9\%} & \multirow{3}{*}{43.9\%} \\
						& IS               &           & 90.80\%           & \textbf{0.93\%}         &                       &                       \\
						& ID               &           & 90.46\%           & 1.27\%           &                       &                       \\ \midrule
						\multirow{6}{*}{ResNet-32}  & OS               & \multirow{3}{*}{92.63\%}           & 92.21\%           & 0.42\%           & \multirow{3}{*}{46.8\%} & \multirow{3}{*}{40.3\%} \\
						& IS               &            & 92.64\%           & \textbf{-0.01\%}           &                       &                       \\
						& ID               &            & 92.33\%           & 0.30\%  &                       &                       \\ \cmidrule(l){2-7} 
						& OS               & 
						\multirow{3}{*}{92.63\%}           & 91.57\%           & 1.06\%           & \multirow{3}{*}{58.0\%} & \multirow{3}{*}{43.9\%} \\
						& IS               &           & 91.92\%           & \textbf{0.71\%}  &                       &                       \\
						& ID               &           & 91.87\%           & 0.76\%           &                       &                       \\ \midrule
						\multirow{6}{*}{ResNet-56}  & OS               & \multirow{3}{*}{93.39\%}           & 93.12\%           & 0.27\%           & \multirow{3}{*}{45.5\%} & \multirow{3}{*}{40.3\%} \\
						& IS               &            & 93.60\%           & -0.21\%          &                       &                       \\
						& ID               &            & 93.66\%           & \textbf{-0.27\%} &                       &                       \\ \cmidrule(l){2-7} 
						& OS               & 
						\multirow{3}{*}{93.39\%}           & 93.03\%           & 0.36\%           & \multirow{3}{*}{58.1\%} & \multirow{3}{*}{43.9\%} \\
						& IS               &            & 93.11\%           & \textbf{0.28\%}  &                       &                       \\
						& ID               &            & 92.86\%           & 0.53\%           &                       &                       \\ \midrule
						\multirow{6}{*}{ResNet-110} & OS               & \multirow{3}{*}{93.68\%}           & 94.01\%           & -0.33\%          & \multirow{3}{*}{45.6\%} & \multirow{3}{*}{40.4\%} \\
						& IS               &            & 94.56\%           & \textbf{-0.88\%} &                       &                       \\
						& ID               &            & 94.28\%           & -0.60\%          &                       &                       \\ \cmidrule(l){2-7} 
						& OS               & 
						\multirow{3}{*}{93.68\%}           & 93.62\%           & 0.06\%           & \multirow{3}{*}{58.1\%} & \multirow{3}{*}{44.0\%} \\
						& IS               &            & 94.16\%           & -0.48\%          &                       &                       \\
						& ID               &            & 94.20\%           & \textbf{-0.52\%} &                       &                       \\ \midrule
						\multirow{6}{*}{DenseNet-40}  & OS               & \multirow{3}{*}{94.22\%}           & 92.71\%           & 1.51\%           & \multirow{3}{*}{44.4\%} & \multirow{3}{*}{60.8\%} \\
						& IS               &            & 93.56\%           & \textbf{0.66\%}          &                       &                       \\
						& ID               &            & 93.34\%           & 0.88\% &                       &                       \\ \cmidrule(l){2-7} 
						& OS               & 
						\multirow{3}{*}{94.22\%}           & 91.58\%           & 2.64\%           & \multirow{3}{*}{59.6\%} & \multirow{3}{*}{68.6\%} \\
						& IS               &            & 92.54\%           & 1.68\%  &                       &                       \\
						& ID               &            & 92.92\%           & \textbf{1.30\%}           &                       &                       \\
						\bottomrule  
				\end{tabular}}
			}
		\end{center}
		\label{tab:diff_prune_schedules}
	\end{table*}
	
	\subsection{Performance w.r.t Pruning Rates Settings}
	\textbf{A constant pruning rate for all layers:} Structured pruning algorithms frequently adopt a constant percentage to discard the filters or channels in each layer. In contrast, we propose a new perspective on how to assign pruning rates for various convolutional layers via their information concentration. Tab. \ref{tab:re_constant_pruning_rate} lists the comparison of the performance where the layer-wise pruning ratios are constant \vs set by our method via information concentration for pruning ResNet-56 on CIFAR-10. The results demonstrate that our method performs better for all the pruning schedules.
	
	\begin{table*}[ht!]
		\caption{Pruning results of the methods where pruning rates are constant \vs set via information concentration for ResNet-56.}
		\begin{center}
			\scalebox{1}{
				\setlength{\tabcolsep}{3.5mm}{
					\begin{tabular}{@{} c|c|cc|c|cc@{}}
						\toprule
						Pruning Schedule                   & Setting  & Baseline & Accuracy & Acc. Drop    & FLOPs Drop  & Params. Drop  \\ \midrule
						\multirow{2}{*}{One-shot}          & Ours & \multirow{2}{*}{93.39\%}           & 93.15\%           & \textbf{0.24\%}  & 45.5\% & 47.4\% \\
						& Constant        &            & 92.85\%           & 0.54\%           & 45.4\% & 46.6\%          \\ \midrule
						\multirow{2}{*}{Iterative Static}  & Ours & \multirow{2}{*}{93.39\%}           & 93.52\%           & \textbf{-0.13\%} & 45.5\% & 47.4\% \\
						& Constant        &            & 93.38\%           & 0.01\%           & 45.4\% & 46.6\%          \\ \midrule
						\multirow{2}{*}{Iterative Dynamic} & Ours & \multirow{2}{*}{93.39\%}           & 93.72\%           & \textbf{-0.33\%} & 45.5\% & 47.4\% \\
						& Constant       &            & 93.24\%           & 0.15\%           & 45.4\% & 46.6\%          \\ 
						\bottomrule
				\end{tabular}}
			}
		\end{center}
		\label{tab:re_constant_pruning_rate}
	\end{table*}
	
	\textbf{Different pruning rates for each layer:} HRank \cite{hrank} empirically determines the pruning rates for the layers, and PFEC \cite{pfec} sets varying numbers of channels to prune each convolutional layer depending on the layer sensitivity. In contrast to them, we compare our method with PFEC and HRank for the purpose of determining the layer-wise pruning ratios via information concentration. In contrast to PFEC and HRank, which prune the network using one-shot pruning schedules and iterative static pruning schedules, respectively, we prune the network with these two paradigms to compare their performance. Tab. \ref{tab:empirical} shows that the accuracy increase achieved by our method exceeds PFEC (0.55\% \vs 0.02\%), and the accuracy drop achieved by our method is lower than HRank (0.00\% \vs 0.09\%) when pruning ResNet-56 on CIFAR-10. The results show that the feasibility of our method can provide theoretical guidance for the settings of layer-wise pruning ratios.
	
	\begin{table*}[ht!]
		\caption{Pruning results by PFEC, HRank and our method for ResNet-56.}
		\begin{center}
			\scalebox{1}{
				\setlength{\tabcolsep}{3.5mm}{
					\begin{tabular}{@{} c|c|cc|c|cc@{}}
						\toprule
						Pruning Schedule                   & Algorithm  & Baseline & Accuracy & Acc. Drop    & FLOPs Drop  & Params. Drop  \\ \midrule
						\multirow{2}{*}{One-shot}          & Ours & 93.39\%           & 93.94\%           & \textbf{-0.55\%}  & 27.9\% & 28.5\% \\
						& PFEC \cite{pfec}          & 93.04\%           & 93.06\%           & -0.02\%           & 27.6\% & 13.7\%          \\ \midrule
						\multirow{2}{*}{Iterative Static}  & Ours & 93.39\%           & 93.39\%           & \textbf{0.00\%} & 50.6\% & 52.2\% \\
						& HRank \cite{hrank}       & 93.26\%           & 93.17\%           & 0.09\%           & 50.0\% & 42.4\% \\
						\bottomrule
				\end{tabular}}
			}
		\end{center}
		\label{tab:empirical}
	\end{table*}
	
	\subsection{Performance w.r.t Pruning Criteria} Tab. \ref{tab:pruning_criteria_cifar} lists the performance achieved by the pruned models using rank \cite{hrank}, entropy \cite{entropy} and Shapley values leveraged as the importance scoring metric for pruning ResNet-56 on CIFAR-10. Among them, Shapley values outperform the others, showing the superiority of Shapley values as the pruning criteria.
	
	\begin{table}[ht!]
		\centering
		\caption{Pruning results \wrt pruning criteria for ResNet-56.}
		\label{tab:pruning_criteria_cifar}
		
		\begin{tabular}{c|c|c}
			\toprule
			Architecture & Pruning Criteria & Accuracy \\
			\midrule
			\multirow{3}{*}{ResNet-56}  
			& Rank & 92.55\% \\
			& Entropy & 91.17\% \\
			& Shapley values & \textbf{92.60\%} \\
			\bottomrule
		\end{tabular}
		
	\end{table} Considering our method adopts a progressive pruning paradigm the same as SFP \cite{sfp}, we compare the performance where the importance scores are $\ell_2$-norm used by SFP and Shapley values leveraged by our method, as Tab. \ref{tab:pruning_criteria_coco} shown. When pruning RetinaNet, FSAF, ATSS and PAA on COCO2017, Shapley values all outperform $\ell_2$-norm. Therefore, in the progressive pruning paradigm, Shapley values shows the superiority to $\ell_2$-norm for evaluating the importance of channels.
	
	\begin{table}[ht!]
		\centering
		\caption{Pruning results \wrt pruning criteria for RetinaNet, FSAF, ATSS and PAA.}
		\label{tab:pruning_criteria_coco}
		
		\begin{tabular}{c|c|c}
			\toprule
			Architecture                       & Pruning Criteria             & mAP \\ \midrule
			
			\multirow{2}{*}{RetinaNet}  
			& $\ell_2$-norm                & 35.40\%  \\
			& Shapley values              & \textbf{37.30\%}  \\ \midrule
			
			\multirow{2}{*}{FSAF}  
			& $\ell_2$-norm                & 35.80\%  \\
			& Shapley values              & \textbf{37.60\%}  \\ \midrule
			
			\multirow{2}{*}{ATSS}  
			& $\ell_2$-norm                & 35.30\%  \\
			& Shapley values              & \textbf{38.60\%}  \\ \midrule
			
			\multirow{2}{*}{PAA}  
			& $\ell_2$-norm                & 37.70\%  \\
			& Shapley values              & \textbf{39.70\%}  \\
			
			\bottomrule
			
		\end{tabular}
		
	\end{table}

	\subsection{Layers Compression Magnitude} 
\begin{figure*}[t!]
		\centering
		
		\subfloat[RetinaNet - FLOPs]{
			\begin{minipage}[t]{0.3\textwidth}
				\centering
				\includegraphics[scale=0.3]{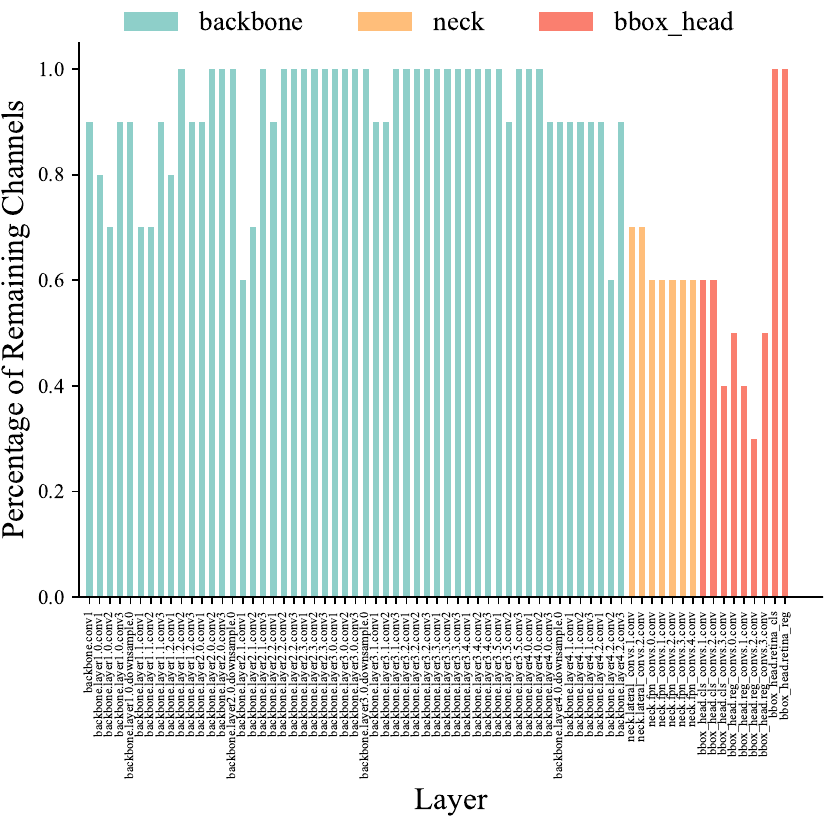}
				\label{fig:remaining_channels_retinanet_flops}
			\end{minipage}
		}
		\subfloat[RetinaNet - Parameters]{
			\begin{minipage}[t]{0.3\textwidth}
				\centering
				\includegraphics[scale=0.3]{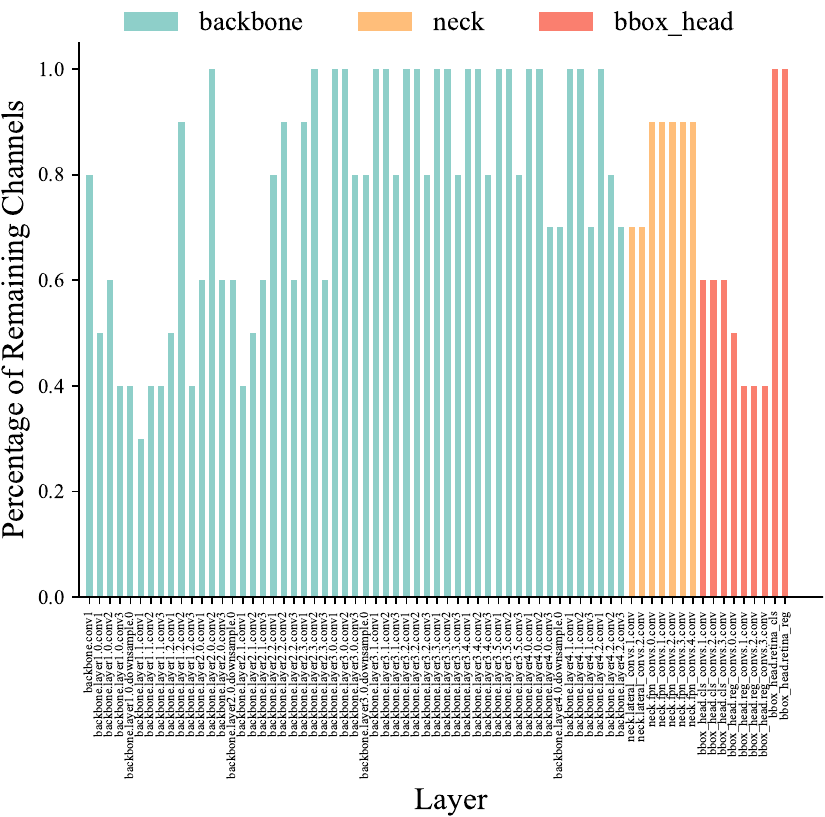}
				\label{fig:remaining_channels_retinanet_parameters}
			\end{minipage}
		}
		\subfloat[RetinaNet - None]{	
			\begin{minipage}[t]{0.3\textwidth}
				\centering
				\includegraphics[scale=0.3]{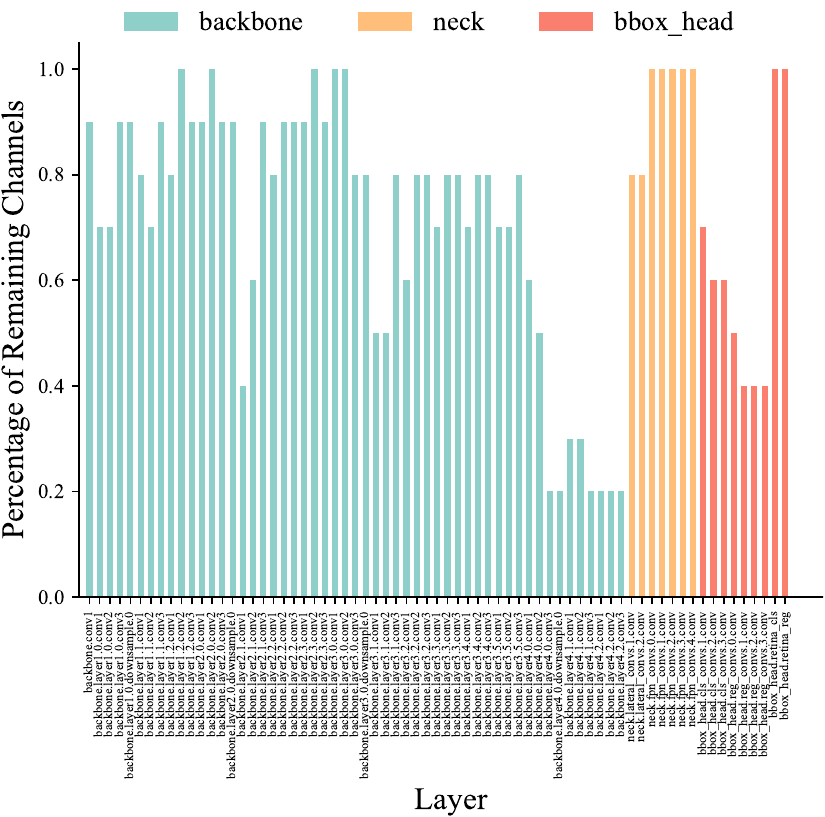}
				\label{fig:remaining_channels_retinanet_none}
			\end{minipage}
		}
		
		\subfloat[FSAF - FLOPs]{
			\begin{minipage}[t]{0.3\textwidth}
				\centering
				\includegraphics[scale=0.3]{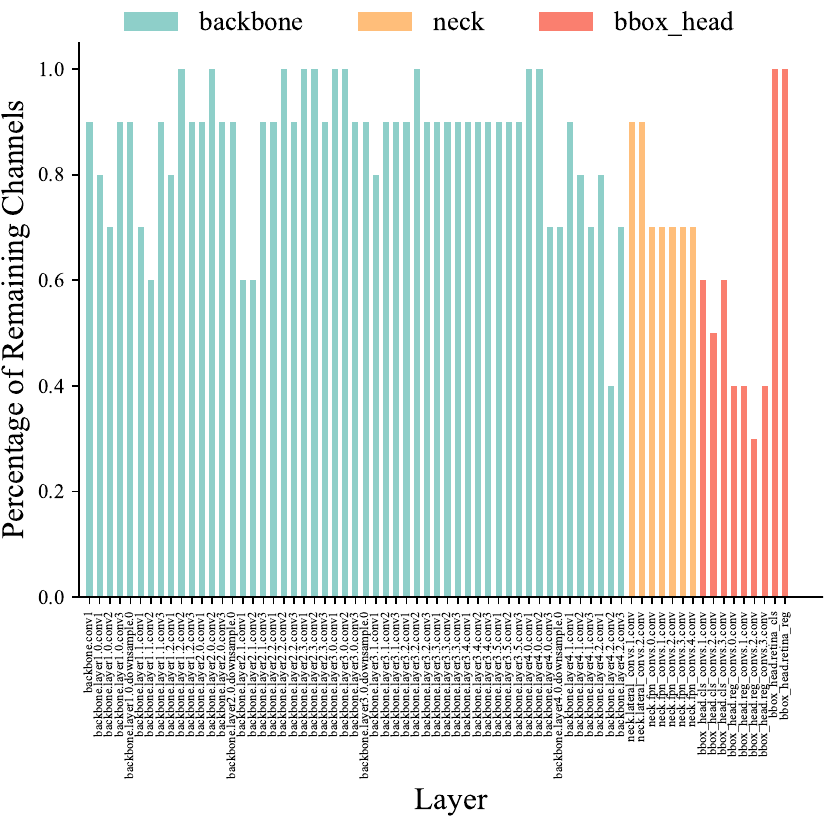}
				\label{fig:remaining_channels_fsaf_flops}
			\end{minipage}
		}
		\subfloat[FSAF - Parameters]{
			\begin{minipage}[t]{0.3\textwidth}
				\centering
				\includegraphics[scale=0.3]{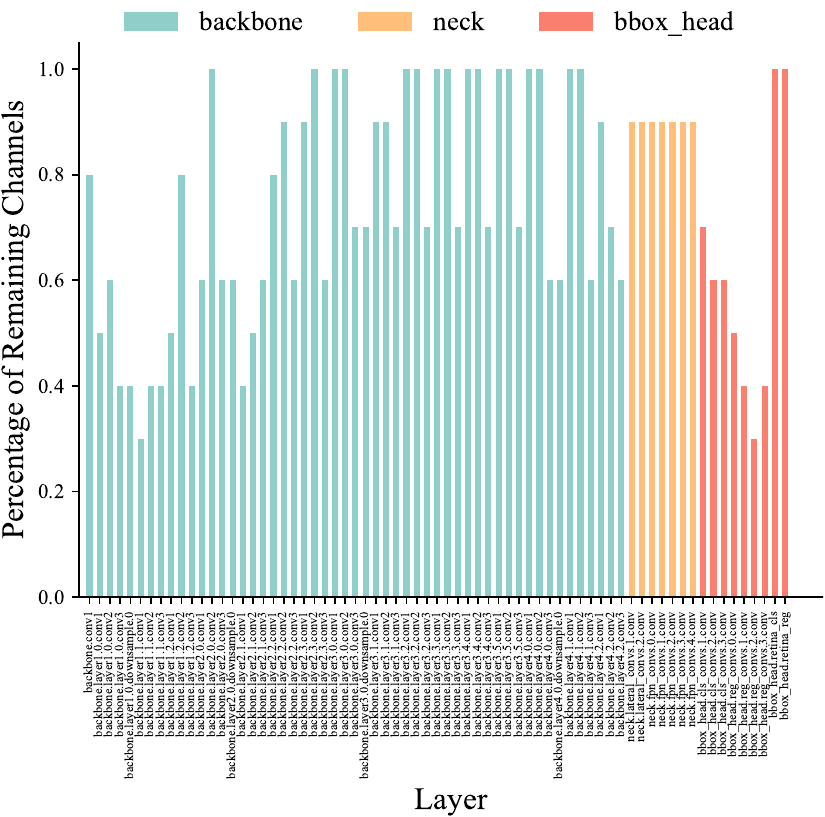}
				\label{fig:remaining_channels_fsaf_parameters}
			\end{minipage}
		}
		\subfloat[FSAF - None]{	
			\begin{minipage}[t]{0.3\textwidth}
				\centering
				\includegraphics[scale=0.3]{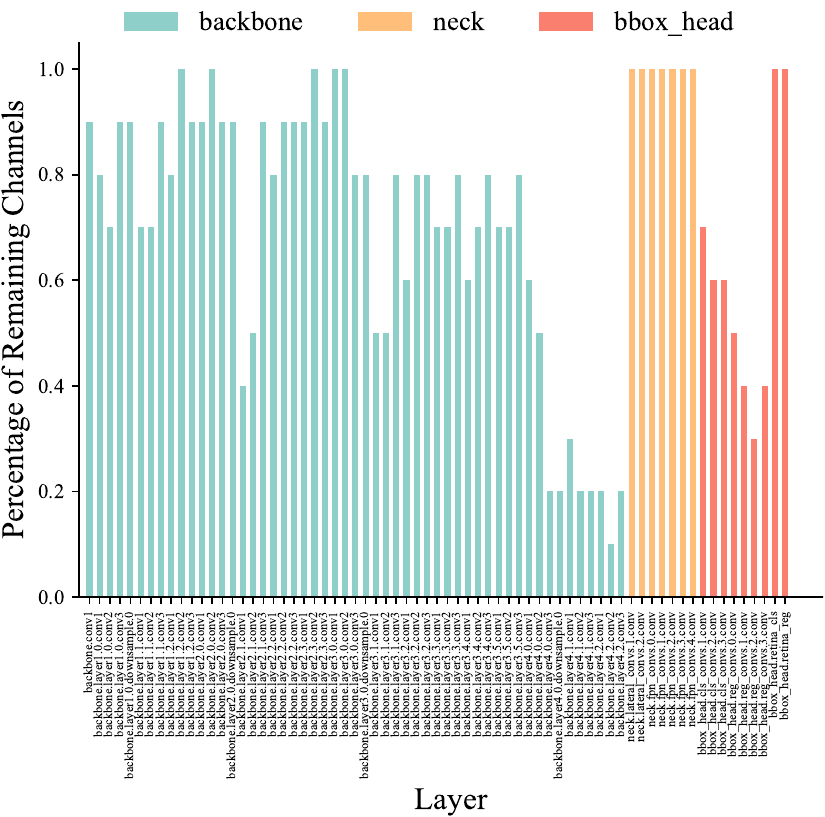}
				\label{fig:remaining_channels_fsaf_none}
			\end{minipage}
		}
		
		\subfloat[ATSS - FLOPs]{
			\begin{minipage}[t]{0.3\textwidth}
				\centering
				\includegraphics[scale=0.3]{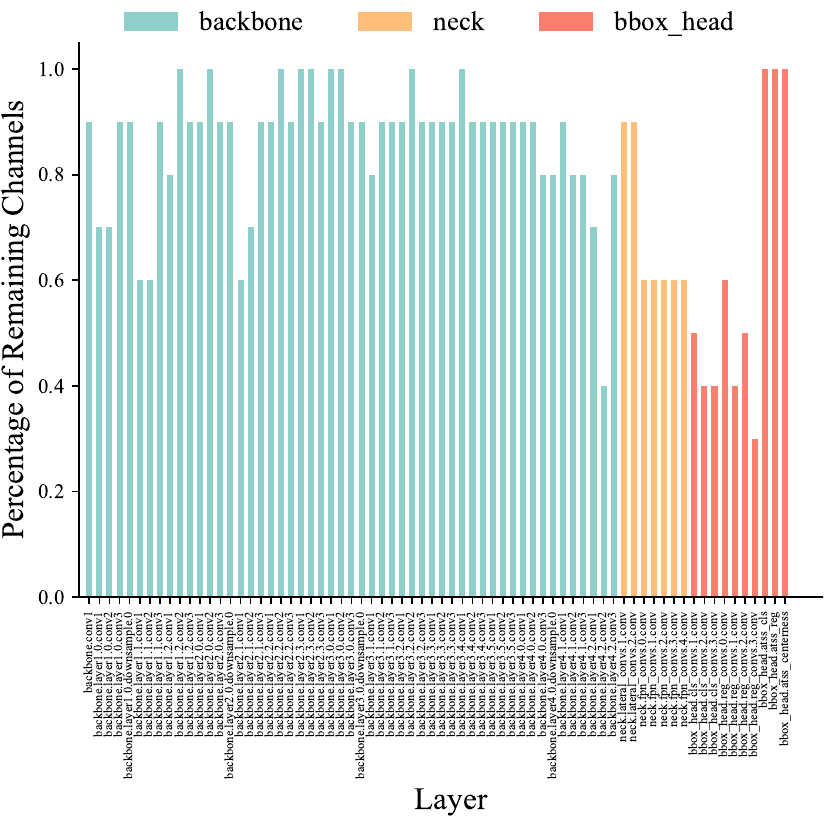}
				\label{fig:remaining_channels_atss_flops}
			\end{minipage}
		}
		\subfloat[ATSS - Parameters]{
			\begin{minipage}[t]{0.3\textwidth}
				\centering
				\includegraphics[scale=0.3]{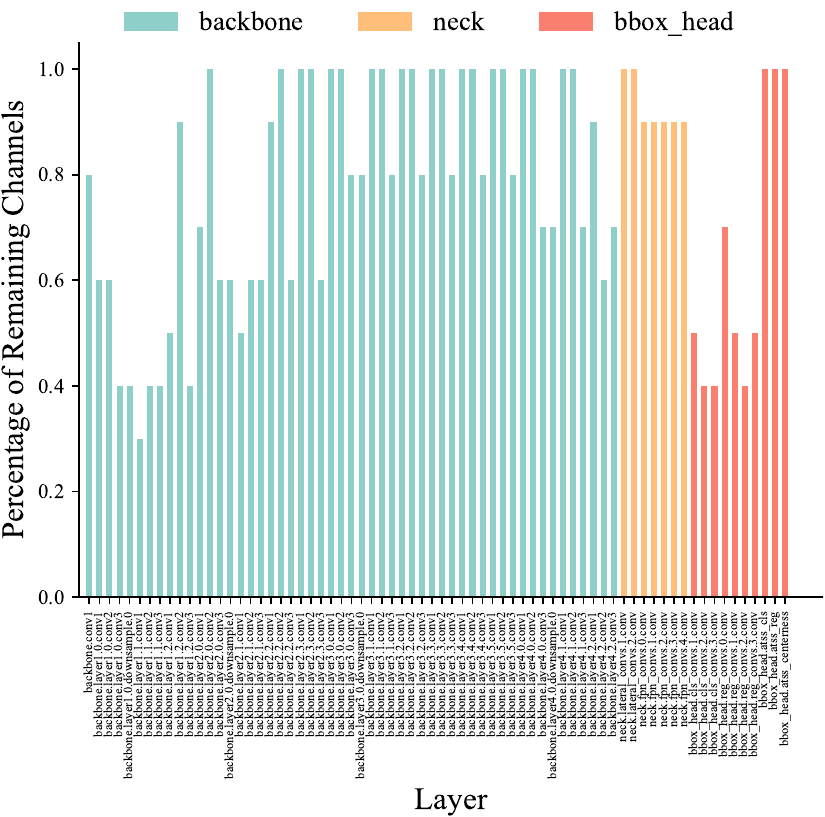}
				\label{fig:remaining_channels_atss_parameters}
			\end{minipage}
		}
		\subfloat[ATSS - None]{	
			\begin{minipage}[t]{0.3\textwidth}
				\centering
				\includegraphics[scale=0.3]{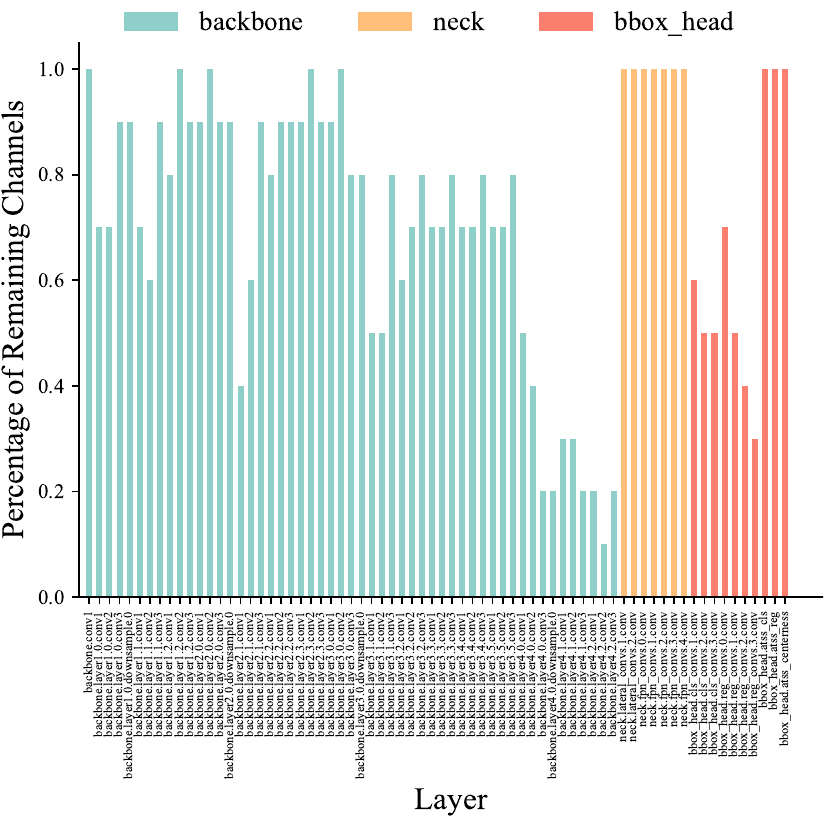}
				\label{fig:remaining_channels_atss_none}
			\end{minipage}
		}
		
		\subfloat[PAA - FLOPs]{
			\begin{minipage}[t]{0.3\textwidth}
				\centering
				\includegraphics[scale=0.3]{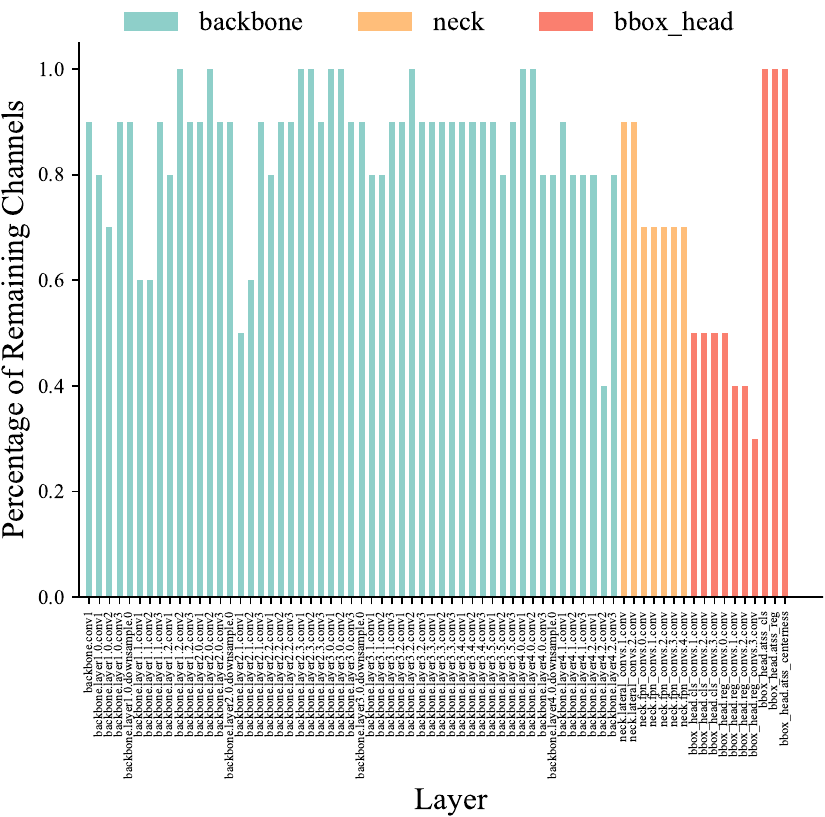}
				\label{fig:remaining_channels_paa_flops}
			\end{minipage}
		}
		\subfloat[PAA - Parameters]{
			\begin{minipage}[t]{0.3\textwidth}
				\centering
				\includegraphics[scale=0.3]{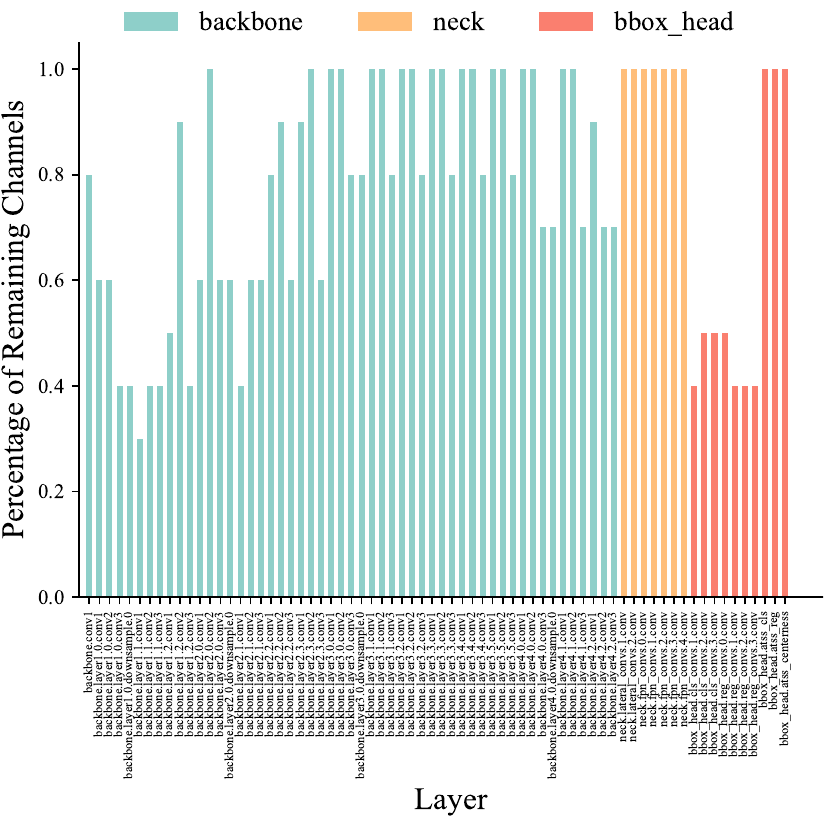}
				\label{fig:remaining_channels_paa_parameters}
			\end{minipage}
		}
		\subfloat[PAA - None]{	
			\begin{minipage}[t]{0.3\textwidth}
				\centering
				\includegraphics[scale=0.3]{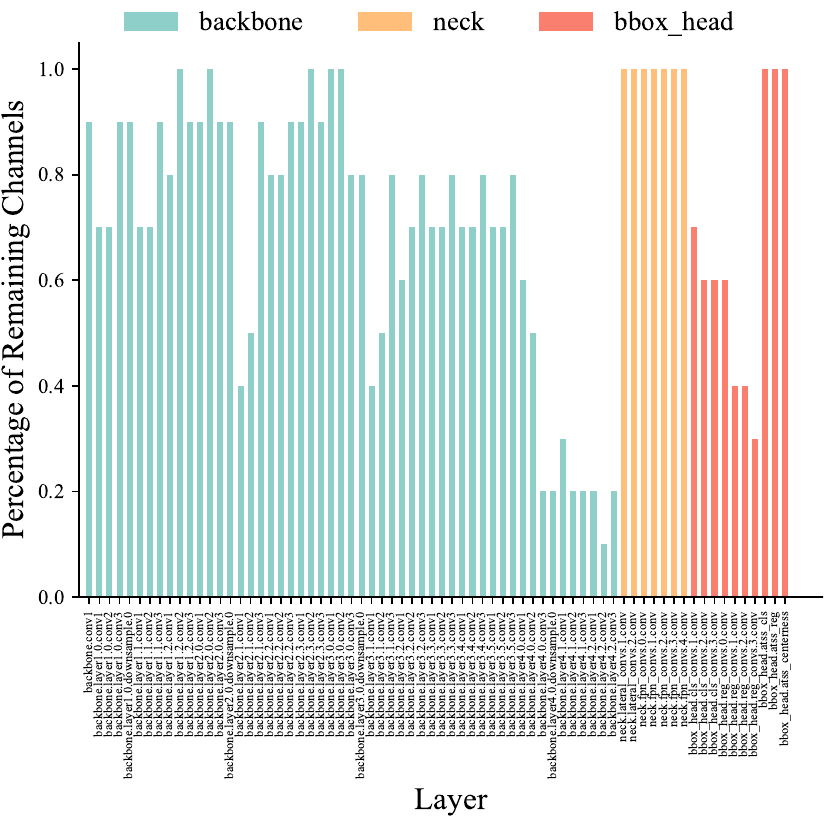}
				\label{fig:remaining_channels_paa_none}
			\end{minipage}
		}
		
		\caption{The percentage of remaining channels in each convolutional layer for RetinaNet, FSAF, ATSS and PAA.}
		\label{fig:remaining_channels}
	\end{figure*}

To demonstrate the compression magnitude of each layer, we visualize the percentages of remaining channels in the convolutional layers of RetinaNet, FSAF, ATSS and PAA in COCO2017, as shown in Figs. \ref{fig:remaining_channels_retinanet_flops} $\sim$ \ref{fig:remaining_channels_paa_none} where the suffix of the backbone denotes the type of normalization for the importance scores. The results show that in the backbones of the pruned models with normalization of importance scores, the former layers remain generally fewer channels than the latter ones. This indicates that the former layers are less informative than the latter ones, which is different from the pruned networks in image classification as Sec. \ref{sec:concentration} demonstrated. In contrast, the latter layers are less important than the former ones in the backbones of the pruned models without normalization of importance scores, leading to worse performance than the backbones with normalization. In the necks of pruned models with normalization, the proportions of remaining channels are lower than those in the backbones, which indicates the backbones play a more important role than the necks so as to achieve excellent performance. Additionally, the lower percentages of remaining channels in the bounding box heads demonstrate that the head is less important than the backbone or neck.
	
	\begin{table*}[ht!]
		\centering
		
		\captionsetup{labelfont={color=red}}
		
		\caption{Accuracy w.r.t. flops for ResNet-110 on CIFAR-10.}
		
		\scalebox{0.95}{
			\begin{tabular}{c|c|c|c|c|c|c|c|c|c|c}
				\hline
				\diagbox[width=10em]{Method}{Accuracy}{FLOPs} 
				& 255.5M& 230.9M& 204.4M (30\%) & 178.8M & 153.3M (50\%) & 127.7M& 102.2M (70\%) & 78.8M & 51.1M (90\%) & 25.5M\\
				\hline
				Ours & 93.75\% & 94.01\% & 93.85\% & 93.80\% & 93.78\% & 93.52\% & 93.51\% & 92.55\% & 90.82\% & 87.05\% \\
				\hline
				FPGM\cite{fpgm} & 93.75\% & 94.00\% & 93.70\% & 94.03\% & 93.50\% & 93.53\% & 92.50\% & 92.51\% & 91.82\% & 88.75\% \\
				\hline
				SFP\cite{sfp} & 93.75\% & 93.90\% & 93.92\% & 93.02\% & 92.50\% & 92.01\% & 91.80\% & 89.82\% & 89.20\% & 84.76\% \\
				\hline
			\end{tabular}
		}
		\label{tab:trade-off}
	\end{table*}
	
	{\textit{E. Performance w.r.t Pruning Rates} To explore the balance between model compression (reducing FLOPs and parameters) and model performance (accuracy or mAP), we performed a new experiment, using the ResNet-101 model on the CIFAR-10 dataset. The result is shown as Tab. \ref{tab:trade-off}. In this experiment, we compared our method with FPGM and SFP at different pruning rates (from 0\% to 90\%). We found that when the pruning rate is less than 50\%, our method did not result in a decrease in model accuracy, and in some cases, the accuracy even improved. However, when the pruning rate exceeded 50\%, the accuracy of the model began to decline. Therefore, we believe that 50\% is a critical turning point in the balance between model compression and performance. Before this point, the impact of model compression on model performance is relatively small, and it may even optimize the model to some extent. However, when the pruning rate exceeds this turning point, model compression begins to have a significant impact on model performance. It is worth noting that we realize the position of this ``turning point'' may vary with different models and datasets. Therefore, our experimental results can only provide a specific example to demonstrate the balance between model compression and performance. In different application scenarios and conditions, the specific ``turning point'' may vary, which calls for further research and exploration of specific models and datasets.}
	
	\section{Conclusion} We present a novel method in channel pruning for reducing the computation overhead and memory footprint. According to information theory, we leverage rank and entropy to indicate the information redundancy and amount for the convolutional layers. Then we obtain an overall indicator as information concentration by combining these two indicators, which reaches a compromise of the information indicated by them. Pruning by Shapley values provides more reliability in channel pruning. In the one-shot and iterative pruning paradigms, we use the information concentration to appropriately pre-define the layer-wise pruning ratios. Besides, we employ Shapley values, which are a potent tool in the interpretability of neural networks, to evaluate the importance of channels in a CNN-based model and discard the least important ones for the purpose of obtaining a compact model. Extensive experiments for pruning architectures in image classification and object detection verify the effectiveness and demonstrate the promising performance of our method. In future work, we will extend our method to more Transformer-based models in not only computer vision but also natural language processing, for the purpose of providing a general pruning framework in network compression.
	
	\bibliographystyle{IEEEtran}
	\bibliography{references}

\begin{thebibliography}{10}
\providecommand{\url}[1]{#1}
\csname url@samestyle\endcsname
\providecommand{\newblock}{\relax}
\providecommand{\bibinfo}[2]{#2}
\providecommand{\BIBentrySTDinterwordspacing}{\spaceskip=0pt\relax}
\providecommand{\BIBentryALTinterwordstretchfactor}{4}
\providecommand{\BIBentryALTinterwordspacing}{\spaceskip=\fontdimen2\font plus
\BIBentryALTinterwordstretchfactor\fontdimen3\font minus
  \fontdimen4\font\relax}
\providecommand{\BIBforeignlanguage}[2]{{%
\expandafter\ifx\csname l@#1\endcsname\relax
\typeout{** WARNING: IEEEtran.bst: No hyphenation pattern has been}%
\typeout{** loaded for the language `#1'. Using the pattern for}%
\typeout{** the default language instead.}%
\else
\language=\csname l@#1\endcsname
\fi
#2}}
\providecommand{\BIBdecl}{\relax}
\BIBdecl

\bibitem{alexnet}
A.~Krizhevsky, I.~Sutskever, and G.~E. Hinton, ``Imagenet classification with
  deep convolutional neural networks,'' \emph{Advances in neural information
  processing systems}, vol.~25, 2012.

\bibitem{vgg}
K.~Simonyan and A.~Zisserman, ``Very deep convolutional networks for
  large-scale image recognition,'' \emph{arXiv preprint arXiv:1409.1556}, 2014.

\bibitem{googlenet}
C.~Szegedy, W.~Liu, Y.~Jia, P.~Sermanet, S.~Reed, D.~Anguelov, D.~Erhan,
  V.~Vanhoucke, and A.~Rabinovich, ``Going deeper with convolutions,'' in
  \emph{Proceedings of the IEEE conference on computer vision and pattern
  recognition}, 2015, pp. 1--9.

\bibitem{resnet}
K.~He, X.~Zhang, S.~Ren, and J.~Sun, ``Deep residual learning for image
  recognition,'' in \emph{Proceedings of the IEEE conference on computer vision
  and pattern recognition}, 2016, pp. 770--778.

\bibitem{faster-rcnn}
S.~Ren, K.~He, R.~Girshick, and J.~Sun, ``Faster r-cnn: Towards real-time
  object detection with region proposal networks,'' \emph{Advances in neural
  information processing systems}, vol.~28, 2015.

\bibitem{yolo}
J.~Redmon, S.~Divvala, R.~Girshick, and A.~Farhadi, ``You only look once:
  Unified, real-time object detection,'' in \emph{Proceedings of the IEEE
  conference on computer vision and pattern recognition}, 2016, pp. 779--788.

\bibitem{ssd}
W.~Liu, D.~Anguelov, D.~Erhan, C.~Szegedy, S.~Reed, C.-Y. Fu, and A.~C. Berg,
  ``Ssd: Single shot multibox detector,'' in \emph{European conference on
  computer vision}.\hskip 1em plus 0.5em minus 0.4em\relax Springer, 2016, pp.
  21--37.

\bibitem{fpn}
T.-Y. Lin, P.~Doll{\'a}r, R.~Girshick, K.~He, B.~Hariharan, and S.~Belongie,
  ``Feature pyramid networks for object detection,'' in \emph{Proceedings of
  the IEEE conference on computer vision and pattern recognition}, 2017, pp.
  2117--2125.

\bibitem{unet}
O.~Ronneberger, P.~Fischer, and T.~Brox, ``U-net: Convolutional networks for
  biomedical image segmentation,'' in \emph{International Conference on Medical
  image computing and computer-assisted intervention}.\hskip 1em plus 0.5em
  minus 0.4em\relax Springer, 2015, pp. 234--241.

\bibitem{fcn}
J.~Long, E.~Shelhamer, and T.~Darrell, ``Fully convolutional networks for
  semantic segmentation,'' in \emph{Proceedings of the IEEE conference on
  computer vision and pattern recognition}, 2015, pp. 3431--3440.

\bibitem{segnet}
V.~Badrinarayanan, A.~Kendall, and R.~Cipolla, ``Segnet: A deep convolutional
  encoder-decoder architecture for image segmentation,'' \emph{IEEE
  transactions on pattern analysis and machine intelligence}, vol.~39, no.~12,
  pp. 2481--2495, 2017.

\bibitem{weight-structured}
S.~J. Kwon, D.~Lee, B.~Kim, P.~Kapoor, B.~Park, and G.-Y. Wei, ``Structured
  compression by weight encryption for unstructured pruning and quantization,''
  in \emph{Proceedings of the IEEE/CVF Conference on Computer Vision and
  Pattern Recognition}, 2020, pp. 1909--1918.

\bibitem{weight-deep}
S.~Han, H.~Mao, and W.~J. Dally, ``Deep compression: Compressing deep neural
  networks with pruning, trained quantization and huffman coding,'' \emph{arXiv
  preprint arXiv:1510.00149}, 2015.

\bibitem{weight-learning}
S.~Han, J.~Pool, J.~Tran, and W.~Dally, ``Learning both weights and connections
  for efficient neural network,'' \emph{Advances in neural information
  processing systems}, vol.~28, 2015.

\bibitem{weight-dynamic}
Y.~Guo, A.~Yao, and Y.~Chen, ``Dynamic network surgery for efficient dnns,''
  \emph{Advances in neural information processing systems}, vol.~29, 2016.

\bibitem{weight-compression}
M.~A. Carreira-Perpin{\'a}n and Y.~Idelbayev, ``"learning-compression"
  algorithms for neural net pruning,'' in \emph{Proceedings of the IEEE
  Conference on Computer Vision and Pattern Recognition}, 2018, pp. 8532--8541.

\bibitem{dynamic}
Y.~Guo, A.~Yao, and Y.~Chen, ``Dynamic network surgery for efficient dnns,''
  \emph{Advances in neural information processing systems}, vol.~29, 2016.

\bibitem{surgeon}
X.~Dong, S.~Chen, and S.~Pan, ``Learning to prune deep neural networks via
  layer-wise optimal brain surgeon,'' \emph{Advances in Neural Information
  Processing Systems}, vol.~30, 2017.

\bibitem{runtime}
J.~Lin, Y.~Rao, J.~Lu, and J.~Zhou, ``Runtime neural pruning,'' \emph{Advances
  in neural information processing systems}, vol.~30, 2017.

\bibitem{pruning}
P.~Molchanov, S.~Tyree, T.~Karras, T.~Aila, and J.~Kautz, ``Pruning
  convolutional neural networks for resource efficient inference,'' \emph{arXiv
  preprint arXiv:1611.06440}, 2016.

\bibitem{variational}
C.~Zhao, B.~Ni, J.~Zhang, Q.~Zhao, W.~Zhang, and Q.~Tian, ``Variational
  convolutional neural network pruning,'' in \emph{Proceedings of the IEEE/CVF
  Conference on Computer Vision and Pattern Recognition}, 2019, pp. 2780--2789.

\bibitem{dais}
Y.~Guan, N.~Liu, P.~Zhao, Z.~Che, K.~Bian, Y.~Wang, and J.~Tang, ``Dais:
  Automatic channel pruning via differentiable annealing indicator search,''
  \emph{IEEE Transactions on Neural Networks and Learning Systems}, pp. 1--12,
  2022.

\bibitem{carrying}
Y.~Zhang, M.~Lin, C.-W. Lin, J.~Chen, Y.~Wu, Y.~Tian, and R.~Ji, ``Carrying out
  cnn channel pruning in a white box,'' \emph{IEEE Transactions on Neural
  Networks and Learning Systems}, 2022.

\bibitem{faster}
J.~Park, S.~Li, W.~Wen, P.~T.~P. Tang, H.~Li, Y.~Chen, and P.~Dubey, ``Faster
  cnns with direct sparse convolutions and guided pruning,'' \emph{arXiv
  preprint arXiv:1608.01409}, 2016.

\bibitem{eie}
S.~Han, X.~Liu, H.~Mao, J.~Pu, A.~Pedram, M.~A. Horowitz, and W.~J. Dally,
  ``Eie: Efficient inference engine on compressed deep neural network,''
  \emph{ACM SIGARCH Computer Architecture News}, vol.~44, no.~3, pp. 243--254,
  2016.

\bibitem{hrank}
M.~Lin, R.~Ji, Y.~Wang, Y.~Zhang, B.~Zhang, Y.~Tian, and L.~Shao, ``Hrank:
  Filter pruning using high-rank feature map,'' in \emph{Proceedings of the
  IEEE/CVF conference on computer vision and pattern recognition}, 2020, pp.
  1529--1538.

\bibitem{cp}
Y.~He, X.~Zhang, and J.~Sun, ``Channel pruning for accelerating very deep
  neural networks,'' in \emph{Proceedings of the IEEE international conference
  on computer vision}, 2017, pp. 1389--1397.

\bibitem{thinet}
J.-H. Luo, J.~Wu, and W.~Lin, ``Thinet: A filter level pruning method for deep
  neural network compression,'' in \emph{Proceedings of the IEEE international
  conference on computer vision}, 2017, pp. 5058--5066.

\bibitem{emerging}
H.~Wang, C.~Qin, Y.~Zhang, and Y.~Fu, ``Emerging paradigms of neural network
  pruning,'' \emph{arXiv preprint arXiv:2103.06460}, 2021.

\bibitem{pfec}
H.~Li, A.~Kadav, I.~Durdanovic, H.~Samet, and H.~P. Graf, ``Pruning filters for
  efficient convnets,'' \emph{arXiv preprint arXiv:1608.08710}, 2016.

\bibitem{sfp}
Y.~He, G.~Kang, X.~Dong, Y.~Fu, and Y.~Yang, ``Soft filter pruning for
  accelerating deep convolutional neural networks,'' \emph{arXiv preprint
  arXiv:1808.06866}, 2018.

\bibitem{fpgm}
Y.~He, P.~Liu, Z.~Wang, Z.~Hu, and Y.~Yang, ``Filter pruning via geometric
  median for deep convolutional neural networks acceleration,'' in
  \emph{Proceedings of the IEEE/CVF Conference on Computer Vision and Pattern
  Recognition}, 2019, pp. 4340--4349.

\bibitem{entropy}
J.-H. Luo and J.~Wu, ``An entropy-based pruning method for cnn compression,''
  \emph{arXiv preprint arXiv:1706.05791}, 2017.

\bibitem{shapley}
\BIBentryALTinterwordspacing
L.~S. Shapley, \emph{17. A Value for n-Person Games}.\hskip 1em plus 0.5em
  minus 0.4em\relax Princeton University Press, 2016, pp. 307--318. [Online].
  Available: \url{https://doi.org/10.1515/9781400881970-018}
\BIBentrySTDinterwordspacing

\bibitem{cicc}
Y.~Chen, Z.~Li, Y.~Yang, L.~Xie, Y.~Liu, L.~Ma, S.~Liu, and G.~Tian, ``Cicc:
  Channel pruning via the concentration of information and contributions of
  channels.'' in \emph{BMVC}, 2022, p. 243.

\bibitem{cifar}
A.~Krizhevsky, G.~Hinton \emph{et~al.}, ``Learning multiple layers of features
  from tiny images,'' 2009.

\bibitem{imagenet}
O.~Russakovsky, J.~Deng, H.~Su, J.~Krause, S.~Satheesh, S.~Ma, Z.~Huang,
  A.~Karpathy, A.~Khosla, M.~Bernstein, A.~Berg, and L.~Fei-Fei, ``Imagenet
  large scale visual recognition challenge,'' \emph{International Journal of
  Computer Vision}, vol. 115, 09 2014.

\bibitem{densenet}
G.~Huang, Z.~Liu, L.~Van Der~Maaten, and K.~Q. Weinberger, ``Densely connected
  convolutional networks,'' in \emph{Proceedings of the IEEE conference on
  computer vision and pattern recognition}, 2017, pp. 4700--4708.

\bibitem{coco}
T.-Y. Lin, M.~Maire, S.~Belongie, J.~Hays, P.~Perona, D.~Ramanan,
  P.~Doll{\'a}r, and C.~L. Zitnick, ``Microsoft coco: Common objects in
  context,'' in \emph{European conference on computer vision}.\hskip 1em plus
  0.5em minus 0.4em\relax Springer, 2014, pp. 740--755.

\bibitem{retinanet}
T.-Y. Lin, P.~Goyal, R.~Girshick, K.~He, and P.~Doll{\'a}r, ``Focal loss for
  dense object detection,'' in \emph{Proceedings of the IEEE international
  conference on computer vision}, 2017, pp. 2980--2988.

\bibitem{fsaf}
C.~Zhu, Y.~He, and M.~Savvides, ``Feature selective anchor-free module for
  single-shot object detection,'' in \emph{Proceedings of the IEEE/CVF
  conference on computer vision and pattern recognition}, 2019, pp. 840--849.

\bibitem{atss}
S.~Zhang, C.~Chi, Y.~Yao, Z.~Lei, and S.~Z. Li, ``Bridging the gap between
  anchor-based and anchor-free detection via adaptive training sample
  selection,'' in \emph{Proceedings of the IEEE/CVF conference on computer
  vision and pattern recognition}, 2020, pp. 9759--9768.

\bibitem{paa}
K.~Kim and H.~S. Lee, ``Probabilistic anchor assignment with iou prediction for
  object detection,'' in \emph{Computer Vision--ECCV 2020: 16th European
  Conference, Glasgow, UK, August 23--28, 2020, Proceedings, Part XXV
  16}.\hskip 1em plus 0.5em minus 0.4em\relax Springer, 2020, pp. 355--371.

\bibitem{non-structured}
Y.~Wang, S.~Ye, Z.~He, X.~Ma, L.~Zhang, S.~Lin, G.~Yuan, S.~Tan, Z.~Li, D.~Fan,
  X.~Qian, X.~Lin, and K.~Ma, ``Non-structured dnn weight pruning considered
  harmful,'' 07 2019.

\bibitem{filter-in-filter}
\BIBentryALTinterwordspacing
F.~Meng, H.~Cheng, K.~Li, H.~Luo, X.~Guo, G.~Lu, and X.~Sun, ``Pruning filter
  in filter,'' in \emph{Advances in Neural Information Processing Systems},
  H.~Larochelle, M.~Ranzato, R.~Hadsell, M.~F. Balcan, and H.~Lin, Eds.,
  vol.~33.\hskip 1em plus 0.5em minus 0.4em\relax Curran Associates, Inc.,
  2020, pp. 17\,629--17\,640. [Online]. Available:
  \url{https://proceedings.neurips.cc/paper/2020/file/ccb1d45fb76f7c5a0bf619f979c6cf36-Paper.pdf}
\BIBentrySTDinterwordspacing

\bibitem{taylor}
P.~Molchanov, S.~Tyree, T.~Karras, T.~Aila, and J.~Kautz, ``Pruning
  convolutional neural networks for resource efficient inference,'' \emph{arXiv
  preprint arXiv:1611.06440}, 2016.

\bibitem{apoz}
H.~Hu, R.~Peng, Y.-W. Tai, and C.-K. Tang, ``Network trimming: A data-driven
  neuron pruning approach towards efficient deep architectures,'' \emph{arXiv
  preprint arXiv:1607.03250}, 2016.

\bibitem{fisher}
L.~Liu, S.~Zhang, Z.~Kuang, A.~Zhou, J.-H. Xue, X.~Wang, Y.~Chen, W.~Yang,
  Q.~Liao, and W.~Zhang, ``Group fisher pruning for practical network
  compression,'' in \emph{International Conference on Machine Learning}.\hskip
  1em plus 0.5em minus 0.4em\relax PMLR, 2021, pp. 7021--7032.

\bibitem{fisher-transformer}
W.~Kwon, S.~Kim, M.~W. Mahoney, J.~Hassoun, K.~Keutzer, and A.~Gholami, ``A
  fast post-training pruning framework for transformers,'' \emph{arXiv preprint
  arXiv:2204.09656}, 2022.

\bibitem{chex}
Z.~Hou, M.~Qin, F.~Sun, X.~Ma, K.~Yuan, Y.~Xu, Y.-K. Chen, R.~Jin, Y.~Xie, and
  S.-Y. Kung, ``Chex: channel exploration for cnn model compression,'' in
  \emph{Proceedings of the IEEE/CVF Conference on Computer Vision and Pattern
  Recognition}, 2022, pp. 12\,287--12\,298.

\bibitem{white-box}
Y.~Zhang, M.~Lin, C.-W. Lin, J.~Chen, Y.~Wu, Y.~Tian, and R.~Ji, ``Carrying out
  cnn channel pruning in a white box,'' \emph{IEEE Transactions on Neural
  Networks and Learning Systems}, vol.~34, no.~10, pp. 7946--7955, 2023.

\bibitem{MFP}
Y.~He, P.~Liu, L.~Zhu, and Y.~Yang, ``Filter pruning by switching to
  neighboring cnns with good attributes,'' \emph{IEEE Transactions on Neural
  Networks and Learning Systems}, vol.~34, no.~10, pp. 8044--8056, 2023.

\bibitem{CLR-RNF}
M.~Lin, L.~Cao, Y.~Zhang, L.~Shao, C.-W. Lin, and R.~Ji, ``Pruning networks
  with cross-layer ranking \& k-reciprocal nearest filters,'' \emph{IEEE
  Transactions on Neural Networks and Learning Systems}, pp. 1--10, 2022.

\bibitem{CATRO}
W.~Hu, Z.~Che, N.~Liu, M.~Li, J.~Tang, C.~Zhang, and J.~Wang, ``Catro: Channel
  pruning via class-aware trace ratio optimization,'' \emph{IEEE Transactions
  on Neural Networks and Learning Systems}, pp. 1--13, 2023.

\bibitem{ABP}
G.~Tian, Y.~Sun, Y.~Liu, X.~Zeng, M.~Wang, Y.~Liu, J.~Zhang, and J.~Chen,
  ``Adding before pruning: Sparse filter fusion for deep convolutional neural
  networks via auxiliary attention,'' \emph{IEEE Transactions on Neural
  Networks and Learning Systems}, 2021.

\bibitem{mdp}
J.~Guo, W.~Ouyang, and D.~Xu, ``Multi-dimensional pruning: A unified framework
  for model compression,'' in \emph{2020 IEEE/CVF Conference on Computer Vision
  and Pattern Recognition (CVPR)}, 2020, pp. 1505--1514.

\bibitem{clfip}
\BIBentryALTinterwordspacing
------, ``Channel pruning guided by classification loss and feature
  importance,'' \emph{Proceedings of the AAAI Conference on Artificial
  Intelligence}, vol.~34, no.~07, pp. 10\,885--10\,892, Apr. 2020. [Online].
  Available: \url{https://ojs.aaai.org/index.php/AAAI/article/view/6720}
\BIBentrySTDinterwordspacing

\bibitem{pcp}
J.~Lin, Z.~Ye, and J.~Wang, ``High efficient compression: Model compression
  method based on channel pruning and knowledge distillation,'' in \emph{2023
  Asia-Europe Conference on Electronics, Data Processing and Informatics
  (ACEDPI)}, 2023, pp. 267--270.

\bibitem{jp}
J.~Guo, J.~Liu, and D.~Xu, ``Jointpruning: Pruning networks along multiple
  dimensions for efficient point cloud processing,'' \emph{IEEE Transactions on
  Circuits and Systems for Video Technology}, vol.~32, no.~6, pp. 3659--3672,
  2022.

\bibitem{3d-p}
------, ``3d-pruning: A model compression framework for efficient 3d action
  recognition,'' \emph{IEEE Transactions on Circuits and Systems for Video
  Technology}, vol.~32, no.~12, pp. 8717--8729, 2022.

\bibitem{actd}
W.~Wang, M.~Chen, S.~Zhao, L.~Chen, J.~Hu, H.~Liu, D.~Cai, X.~He, and W.~Liu,
  ``Accelerate cnns from three dimensions: A comprehensive pruning framework,''
  in \emph{International Conference on Machine Learning}.\hskip 1em plus 0.5em
  minus 0.4em\relax PMLR, 2021, pp. 10\,717--10\,726.

\bibitem{nisp}
R.~Yu, A.~Li, C.-F. Chen, J.-H. Lai, V.~I. Morariu, X.~Han, M.~Gao, C.-Y. Lin,
  and L.~S. Davis, ``Nisp: Pruning networks using neuron importance score
  propagation,'' in \emph{Proceedings of the IEEE Conference on Computer Vision
  and Pattern Recognition}, 2018, pp. 9194--9203.

\bibitem{cc}
Y.~Li, S.~Lin, J.~Liu, Q.~Ye, M.~Wang, F.~Chao, F.~Yang, J.~Ma, Q.~Tian, and
  R.~Ji, ``Towards compact cnns via collaborative compression,'' in
  \emph{Proceedings of the IEEE/CVF Conference on Computer Vision and Pattern
  Recognition}, 2021, pp. 6438--6447.

\bibitem{spp}
H.~Wang, Q.~Zhang, Y.~Wang, and H.~Hu, ``Structured probabilistic pruning for
  convolutional neural network acceleration,'' \emph{arXiv preprint
  arXiv:1709.06994}, 2017.

\bibitem{slimming}
Z.~Liu, J.~Li, Z.~Shen, G.~Huang, S.~Yan, and C.~Zhang, ``Learning efficient
  convolutional networks through network slimming,'' in \emph{Proceedings of
  the IEEE international conference on computer vision}, 2017, pp. 2736--2744.

\bibitem{amc}
Y.~He, J.~Lin, Z.~Liu, H.~Wang, L.-J. Li, and S.~Han, ``Amc: Automl for model
  compression and acceleration on mobile devices,'' in \emph{Proceedings of the
  European conference on computer vision (ECCV)}, 2018, pp. 784--800.

\bibitem{ddpg}
T.~P. Lillicrap, J.~J. Hunt, A.~Pritzel, N.~Heess, T.~Erez, Y.~Tassa,
  D.~Silver, and D.~Wierstra, ``Continuous control with deep reinforcement
  learning,'' \emph{arXiv preprint arXiv:1509.02971}, 2015.

\bibitem{npgr}
H.~Wang, C.~Qin, Y.~Zhang, and Y.~Fu, ``Neural pruning via growing
  regularization,'' \emph{arXiv preprint arXiv:2012.09243}, 2020.

\bibitem{gal}
S.~Lin, R.~Ji, C.~Yan, B.~Zhang, L.~Cao, Q.~Ye, F.~Huang, and D.~Doermann,
  ``Towards optimal structured cnn pruning via generative adversarial
  learning,'' in \emph{Proceedings of the IEEE/CVF Conference on Computer
  Vision and Pattern Recognition}, 2019, pp. 2790--2799.

\bibitem{acp}
Z.~Li, Y.~Sun, G.~Tian, L.~Xie, Y.~Liu, H.~Su, and Y.~He, ``A compression
  pipeline for one-stage object detection model,'' \emph{Journal of Real-Time
  Image Processing}, vol.~18, no.~6, pp. 1949--1962, 2021.

\bibitem{SCL}
Z.~Tang, L.~Luo, B.~Xie, Y.~Zhu, R.~Zhao, L.~Bi, and C.~Lu, ``Automatic sparse
  connectivity learning for neural networks,'' \emph{IEEE Transactions on
  Neural Networks and Learning Systems}, vol.~34, no.~10, pp. 7350--7364, 2023.

\bibitem{information_theory}
D.~J. MacKay, D.~J. Mac~Kay \emph{et~al.}, \emph{Information theory, inference
  and learning algorithms}.\hskip 1em plus 0.5em minus 0.4em\relax Cambridge
  university press, 2003.

\bibitem{captum}
N.~Kokhlikyan, V.~Miglani, M.~Martin, E.~Wang, B.~Alsallakh, J.~Reynolds,
  A.~Melnikov, N.~Kliushkina, C.~Araya, S.~Yan \emph{et~al.}, ``Captum: A
  unified and generic model interpretability library for pytorch,'' \emph{arXiv
  preprint arXiv:2009.07896}, 2020.

\bibitem{polynomial}
J.~Castro, D.~G{\'o}mez, and J.~Tejada, ``Polynomial calculation of the shapley
  value based on sampling,'' \emph{Computers \& Operations Research}, vol.~36,
  no.~5, pp. 1726--1730, 2009.

\bibitem{pytorch}
A.~Paszke, S.~Gross, S.~Chintala, G.~Chanan, E.~Yang, Z.~DeVito, Z.~Lin,
  A.~Desmaison, L.~Antiga, and A.~Lerer, ``Automatic differentiation in
  pytorch,'' 2017.

\bibitem{sss}
Z.~Huang and N.~Wang, ``Data-driven sparse structure selection for deep neural
  networks,'' in \emph{Proceedings of the European conference on computer
  vision (ECCV)}, 2018, pp. 304--320.

\bibitem{dsa}
X.~Ning, T.~Zhao, W.~Li, P.~Lei, Y.~Wang, and H.~Yang, ``Dsa: More efficient
  budgeted pruning via differentiable sparsity allocation,'' in \emph{European
  Conference on Computer Vision}.\hskip 1em plus 0.5em minus 0.4em\relax
  Springer, 2020, pp. 592--607.

\bibitem{pgmpf}
L.~Cai, Z.~An, C.~Yang, Y.~Yan, and Y.~Xu, ``Prior gradient mask guided
  pruning-aware fine-tuning,'' in \emph{Proceedings of the AAAI Conference on
  Artificial Intelligence}, vol.~36, no.~1, 2022, pp. 140--148.

\bibitem{pscratch}
Y.~Wang, X.~Zhang, L.~Xie, J.~Zhou, H.~Su, B.~Zhang, and X.~Hu, ``Pruning from
  scratch,'' in \emph{Proceedings of the AAAI Conference on Artificial
  Intelligence}, vol.~34, no.~07, 2020, pp. 12\,273--12\,280.

\bibitem{gcnna}
\BIBentryALTinterwordspacing
D.~Jiang, Y.~Cao, and Q.~Yang, ``On the channel pruning using graph convolution
  network for convolutional neural network acceleration,'' in \emph{Proceedings
  of the Thirty-First International Joint Conference on Artificial
  Intelligence, {IJCAI-22}}, L.~D. Raedt, Ed.\hskip 1em plus 0.5em minus
  0.4em\relax International Joint Conferences on Artificial Intelligence
  Organization, 7 2022, pp. 3107--3113, main Track. [Online]. Available:
  \url{https://doi.org/10.24963/ijcai.2022/431}
\BIBentrySTDinterwordspacing

\bibitem{decore}
M.~Alwani, Y.~Wang, and V.~Madhavan, ``Decore: Deep compression with
  reinforcement learning,'' in \emph{Proceedings of the IEEE/CVF Conference on
  Computer Vision and Pattern Recognition}, 2022, pp. 12\,349--12\,359.

\bibitem{dbp}
W.~Wang, S.~Zhao, M.~Chen, J.~Hu, D.~Cai, and H.~Liu, ``Dbp: discrimination
  based block-level pruning for deep model acceleration,'' \emph{arXiv preprint
  arXiv:1912.10178}, 2019.

\bibitem{graph}
Y.~Lu, W.~Yang, Y.~Zhang, J.~Wang, S.~Gong, Z.~Chen, Z.~Chen, Q.~Xuan, and
  X.~Yang, ``Graph modularity: Towards understanding the cross-layer transition
  of feature representations in deep neural networks,'' \emph{arXiv preprint
  arXiv:2111.12485}, 2021.

\bibitem{tpp}
H.~Wang and Y.~Fu, ``Trainability preserving neural structured pruning,''
  \emph{arXiv preprint arXiv:2207.12534}, 2022.

\bibitem{gnnrl}
\BIBentryALTinterwordspacing
S.~Yu, A.~Mazaheri, and A.~Jannesari, ``Topology-aware network pruning using
  multi-stage graph embedding and reinforcement learning,'' in
  \emph{Proceedings of the 39th International Conference on Machine Learning},
  ser. Proceedings of Machine Learning Research, K.~Chaudhuri, S.~Jegelka,
  L.~Song, C.~Szepesvari, G.~Niu, and S.~Sabato, Eds., vol. 162.\hskip 1em plus
  0.5em minus 0.4em\relax PMLR, 17--23 Jul 2022, pp. 25\,656--25\,667.
  [Online]. Available: \url{https://proceedings.mlr.press/v162/yu22e.html}
\BIBentrySTDinterwordspacing

\bibitem{rethinking}
Z.~Liu, M.~Sun, T.~Zhou, G.~Huang, and T.~Darrell, ``Rethinking the value of
  network pruning,'' \emph{arXiv preprint arXiv:1810.05270}, 2018.

\bibitem{mmdetection}
K.~Chen, J.~Wang, J.~Pang, Y.~Cao, Y.~Xiong, X.~Li, S.~Sun, W.~Feng, Z.~Liu,
  J.~Xu \emph{et~al.}, ``Mmdetection: Open mmlab detection toolbox and
  benchmark,'' \emph{arXiv preprint arXiv:1906.07155}, 2019.

\bibitem{deit}
H.~Touvron, M.~Cord, M.~Douze, F.~Massa, A.~Sablayrolles, and H.~J{\'e}gou,
  ``Training data-efficient image transformers \& distillation through
  attention,'' in \emph{International conference on machine learning}.\hskip
  1em plus 0.5em minus 0.4em\relax PMLR, 2021, pp. 10\,347--10\,357.

\bibitem{vit}
A.~Dosovitskiy, L.~Beyer, A.~Kolesnikov, D.~Weissenborn, X.~Zhai,
  T.~Unterthiner, M.~Dehghani, M.~Minderer, G.~Heigold, S.~Gelly \emph{et~al.},
  ``An image is worth 16x16 words: Transformers for image recognition at
  scale,'' \emph{arXiv preprint arXiv:2010.11929}, 2020.

\bibitem{bert}
J.~Devlin, M.-W. Chang, K.~Lee, and K.~Toutanova, ``Bert: Pre-training of deep
  bidirectional transformers for language understanding,'' \emph{arXiv preprint
  arXiv:1810.04805}, 2018.

\bibitem{mnli}
A.~Williams, N.~Nangia, and S.~R. Bowman, ``A broad-coverage challenge corpus
  for sentence understanding through inference,'' \emph{arXiv preprint
  arXiv:1704.05426}, 2017.

\bibitem{unet++}
Z.~Zhou, M.~M. Rahman~Siddiquee, N.~Tajbakhsh, and J.~Liang, ``Unet++: A nested
  u-net architecture for medical image segmentation,'' in \emph{Deep Learning
  in Medical Image Analysis and Multimodal Learning for Clinical Decision
  Support: 4th International Workshop, DLMIA 2018, and 8th International
  Workshop, ML-CDS 2018, Held in Conjunction with MICCAI 2018, Granada, Spain,
  September 20, 2018, Proceedings 4}.\hskip 1em plus 0.5em minus 0.4em\relax
  Springer, 2018, pp. 3--11.

\bibitem{busi}
W.~Al-Dhabyani, M.~Gomaa, H.~Khaled, and A.~Fahmy, ``Dataset of breast
  ultrasound images,'' \emph{Data in brief}, vol.~28, p. 104863, 2020.

\bibitem{isic}
N.~C. Codella, D.~Gutman, M.~E. Celebi, B.~Helba, M.~A. Marchetti, S.~W. Dusza,
  A.~Kalloo, K.~Liopyris, N.~Mishra, H.~Kittler \emph{et~al.}, ``Skin lesion
  analysis toward melanoma detection: A challenge at the 2017 international
  symposium on biomedical imaging (isbi), hosted by the international skin
  imaging collaboration (isic),'' in \emph{2018 IEEE 15th international
  symposium on biomedical imaging (ISBI 2018)}.\hskip 1em plus 0.5em minus
  0.4em\relax IEEE, 2018, pp. 168--172.

\end{thebibliography}
	
\end{document}